\documentclass[final,3p,times]{elsarticle}

\usepackage{amssymb}
\usepackage{amsthm}
\usepackage{amsmath}
\usepackage{libertine}
\usepackage{libertinust1math}
\usepackage[backref=page]{hyperref}
\usepackage{paralist}
\usepackage{graphicx}
\usepackage{algorithm}
\usepackage{algpseudocode}
\usepackage{xcolor}
\usepackage{subcaption}
\usepackage{makecell}
\usepackage{xspace}
\usepackage{enumitem}
\usepackage{rotating}
\usepackage{array}
\usepackage[T1]{fontenc}
\newcommand{\refcite}[1]{Ref.~\cite{#1}}

\graphicspath{{./figures/}}

\makeatletter
\newcommand\newtag[2]{#1\def\@currentlabel{#1}\label{#2}}
\makeatother

\algnewcommand{\Inputs}[1]{%
  \State \textbf{Inputs:}
  \Statex \hspace*{\algorithmicindent}\parbox[t]{.8\linewidth}{\raggedright #1}
}
\algnewcommand{\Initialize}[1]{%
  \State \textbf{Initialize:}
  \Statex \hspace*{\algorithmicindent}\parbox[t]{.8\linewidth}{\raggedright #1}
}

\begin{document}

\begin{frontmatter}



\title{How trust networks shape students' opinions about the proficiency of artificially intelligent assistants}

\author[inst1,inst2]{Yutong Bu\corref{cor1}}
\ead{buy1@student.unimelb.edu.au}
\cortext[cor1]{Corresponding author}
\author[inst2,inst3]{Andrew Melatos}
\author[inst1]{Robin Evans}

\affiliation[inst1]{organization={Electrical and Electronic Engineering Department},
           addressline={University of Melbourne}, 
           city={Parkville},
           postcode={VIC 3010},
           country={Australia}}

\affiliation[inst2]{organization={Australian Research Council Centre of Excellence for Gravitational Wave Discovery (OzGrav)},
           addressline={University of Melbourne}, 
           city={Parkville},
           postcode={VIC 3010},
           country={Australia}}

\affiliation[inst3]{organization={School of Physics},
           addressline={University of Melbourne}, 
           city={Parkville},
           postcode={VIC 3010},
           country={Australia}}

\begin{abstract}
The rising use of educational tools controlled by artificial intelligence (AI) has provoked a debate about their proficiency.
While intrinsic proficiency, especially in tasks such as grading, has been measured and studied extensively, perceived proficiency remains underexplored. 
Here it is shown through Monte Carlo multi-agent simulations that trust networks among students influence their perceptions of the proficiency of an AI tool.
A probabilistic opinion dynamics model is constructed, in which every student's perceptions are described by a probability density function (PDF), which is updated at every time step through independent, personal observations and peer pressure shaped by trust relationships.
It is found that students infer correctly the AI tool's proficiency $\theta_{\rm AI}$ in allies-only networks (i.e.\ high trust networks).
AI-avoiders reach asymptotic learning faster than AI-users, and the asymptotic learning time for AI-users decreases as their number increases. 
However, asymptotic learning is disrupted even by a single partisan, who is stubbornly incorrect in their belief $\theta_{\rm p} \neq \theta_{\rm AI}$, making other students' beliefs vacillate indefinitely between $\theta_{\rm p}$ and $\theta_{\rm AI}$.
In opponents-only (low trust) networks, all students reach asymptotic learning, but only a minority infer $\theta_{\rm AI}$ correctly. 
AI-users have a small advantage over AI-avoiders in reaching the right conclusion.
The outcomes in allies-only and opponents-only networks depend weakly on network size $n$.
In mixed networks, students may exhibit turbulent nonconvergence and intermittency, or achieve asymptotic learning, depending on the relationships between partisans and AI-users.
In smaller mixed networks with $n \lesssim 10$ students, the long-term outcome is affected by whether a partisan teacher is an AI-skeptic ($\theta_{\rm p} < \theta_{\rm AI}$) or an AI-promoter ($\theta_{\rm p} \geq \theta_{\rm AI}$).
In larger mixed networks with $n \gtrsim 10^2$, students are more likely to infer $\theta_{\rm p}$ instead of $\theta_{\rm AI}$. 
The educational implications of the results are discussed briefly in the context of designing robust usage policies for AI tools, with an emphasis on the unintended and inequitable consequences which arise sometimes from counterintuitive network effects.  

\end{abstract}




\end{frontmatter}


\section{Introduction}
\label{sec:intro}

The growing use of artificial intelligence (AI) tools in educational settings is generating debate \cite{crompton_artificial_2023}. 
AI tools are deployed by teachers, e.g.\ to author lesson plans \cite{van_den_berg_chatgpt_2023}, compose feedback \cite{liu_teaching_2024}, and grade assessment tasks \cite{rutner_use_2022}. 
They are also deployed by students, e.g.\ to conduct background research \cite{okonkwo_chatbots_2021,labadze_role_2023} and even generate answers to assessment tasks directly \cite{vuckovic_attitudes_2020}. 
A multifarious conversation is underway about AI-assisted teaching and learning: its ethics and philosophical underpinnings \cite{okonkwo_chatbots_2021,hwang_vision_2020}, its impact on educational goals (e.g.\ liberal, vocational) and hence on society \cite{bharadiya_artificial_2023,khogali_blended_2023}, the practicalities of its use \cite{reiss_use_2021}, the pace and socioeconomic uniformity of its adoption \cite{jeon_speed_2024,sharma_ai_2022,casal-otero_ai_2023}, its fairness and bias \cite{ferrara_fairness_2024}, and so on. 
Important issues of academic integrity arise, especially when students use AI tools to do assessment tasks with zero or minimal human intervention \cite{reiss_use_2021}. Policy frameworks and guardrails are being developed to combat the latter risk \cite{the_european_commission_ethical_nodate,chan_comprehensive_2023}.

One key question about AI-assisted teaching and learning is how well AI tools can grade or respond to assessment tasks respectively, when judged by a human expert. 
Colloquially speaking, how intelligent is the AI? There are two facets to the question. 
First, what is the intrinsic proficiency of the AI? Many studies have been performed to quantify proficiency statistically for objective assessment tasks, e.g.\ mathematical questions with a unique correct answer \cite{zhang_automatic_2022,kortemeyer_toward_2023}, and subjective assessment tasks, e.g.\ writing a cogent nonfiction essay or a piece of creative fiction \cite{vijaya_shetty_essay_2022}. 
The results depend strongly on the specific task and AI tool \cite{hall_identifying_2023} but have exceeded expectations in some quarters, leading to statements such as (verbatim quotes) ``ChatGPT-3.5 performed on average at the level of a C+ student'' \cite{choi_chatgpt_2023} and ``GPT-4 performed at a level comparable to the best performing undergraduate medical student'' \cite{meaney_comparing_2023}.
Beyond intrinsic proficiency, there is a second facet: what is the perceived proficiency of the AI? 
Perceived proficiency has received less attention than intrinsic proficiency, partly because it is harder to measure rigorously, and partly because there is a tacit assumption sometimes that it is correlated with intrinsic proficiency and therefore does not warrant separate treatment. 
In reality, perceived and intrinsic proficiencies often differ. For example, large language models (LLMs) may hallucinate (i.e.\ return false answers) in a superficially plausible manner \cite{ji_survey_2023} and indeed may lie confidently \cite{ribeiro_why_2016}, posing risks for credulous users who overestimate an LLM's intrinsic proficiency \cite{openai_gpt-4_2024}.

Perceived proficiency is crucial because it is the chief determinant of how fast and uniformly AI tools are adopted \cite{wang_factors_2021}.
It is a socially emergent construct.
In this paper, we study how students in a networked cohort form opinions about the proficiency of an AI tool, based on observations of their own performance when they do or do not use the tool, as well as exposure to the opinions of other students in the network. 
A central factor is how the intrinsic performance of the AI tool interacts with trust relationships between students. 
As just one example, student X may develop an inaccurate perception about the AI tool's proficiency (e.g.\ overly optimistic or pessimistic) and transmit it by contagion to other students who trust student X. 
Such phenomena can be investigated through multi-agent simulations and related techniques from the field of opinion dynamics \cite{ye_opinion_2019}, which studies how individual opinions evolve through peer pressure within a community or group. 
Related analyses have been conducted recently to model opinions about media bias (e.g.\ the political bias of a newspaper, as displayed through its editorials) within a network of political allies and opponents \cite{low_discerning_2022,low_vacillating_2022,bu_discerning_2023,deffuant_mixing_2000,degroot_reaching_1974,fang_opinion_2020,shi_evolution_2016,martins_continuous_2008}.
Counterintuitive outcomes emerge under certain realistic conditions, e.g.\ agents in the network exhibit unstable opinions (vacillating intermittently, or failing to converge) \cite{low_vacillating_2022}, converge faster on false than true opinions \cite{low_discerning_2022,bu_discerning_2023}, or develop weak or even negative correlations between perceived and intrinsic media biases \cite{bu_discerning_2023}. 
Counterintuitive network outcomes driven by similar dynamics may also occur in educational settings, when students form opinions about the proficiency of AI assistants. 
It is therefore important to understand under what conditions this may occur when developing policy frameworks and guardrails around the use of AI assistants, lest the adoption of AI tools is disturbed by network effects unrelated to the ethical or technical characteristics of the tools themselves. 

The paper is structured as follows. 
Section \ref{sec:motivation} introduces an idealized opinion dynamics model, in which a cohort of students networked through trust relationships form probabilistic opinions about the proficiency of an AI tool through a combination of personal observation and peer pressure.
Monte Carlo simulations are performed to test how accurately the students infer the intrinsic proficiency of the AI tool, when a subset of the network uses AI to complete assessments, and the rest of the network does not. 
The results of the simulations are analyzed for allies-only, opponents-only, and mixed networks in Section \ref{sec:allies}, where the class or study group is relatively small ($n\lesssim 10$ students), as in some high-school settings or postgraduate university degrees for example.
We quantify the difference in student performance with and without the AI tool, focusing again on counterintuitive or unintended consequences which emerge socially, e.g.\ are there circumstances when most of the cohort underperforms relative to its intrinsic proficiency despite the nominal advantage of AI assistance? 
The biases engendered by partisan students and teachers are studied in Section \ref{sec:partisan_student}. 
The results in Sections \ref{sec:allies} and \ref{sec:partisan_student} are extended to larger networks with $n\gtrsim 10$ students in Section \ref{sec:larger_networks}, with applications to cross-disciplinary service subjects in some undergraduate university degrees for example.
We offer a preliminary discussion of pedagogical and policy implications in Section \ref{sec:Tentative_educational_implementation}, and summarize the conclusions in Section \ref{sec:conclusion}.

\section{Updating beliefs about the proficiency of an AI tool}
\label{sec:motivation}
To fix ideas, let us consider the following highly idealized situation. A class of $n$ students performs a sequence of $T$ assessment tasks. 
Some students perform the assessment tasks themselves, without assistance from an AI tool, and receive a score reflecting their own proficiency. 
Other students delegate the assessment tasks wholly to an AI tool and receive a score reflecting the AI tool's proficiency. 
For the sake of simplicity, the students do not collude when performing the assessments (although they do share their beliefs about the AI tool's proficiency, as discussed below), nor do they change their strategy (spurning or seeking AI assistance) from one assessment task to the next.
There is only one AI tool. 
At every time-step $1 \leq t \leq T$, every student harbors a set of uncertain beliefs about the hidden proficiency of the AI tool, which is parameterized by the scalar $\theta_{\rm AI}$ (with $0\leq \theta_{\rm AI} \leq 1$, where zero means not proficient at all, and one means flawless). 
The beliefs of the $i$-th student at time $t$ about $\theta_{\rm AI}$ are encapsulated in a probability density function (PDF) $x_i(t,\theta)$, which evolves as the sequence of assessment tasks progresses. 
That is, every student's belief about the proficiency of the AI tool changes with time, even though their inclination to spurn or seek AI assistance does not change in the simplest version of the model. 
Refinements which relax these assumptions will be implemented in a forthcoming paper. 

In this section, we define a two-step update rule for $x_i(t,\theta)$, which involves a mix of independent observations and peer pressure. 
In step one, described in Section \ref{subsec:observing_outcomes}, the $i$-th student applies Bayes's theorem to their score $S_i(t)$ in the $t$-th assignment to infer a provisional posterior $x_i'(t+1/2,\theta)$. 
In step two, described in Section \ref{subsec:peer_pressure}, the $i$-th student shares their beliefs with other students in the class through a linear but non-Bayesian mixing rule, converging towards the beliefs of trusted classmates and diverging from the beliefs of mistrusted classmates. 
In Section \ref{subsec:relation_to_existing_models}, the idealizations and approximations in the model are discussed, and the model is related to others dealing with similar educational and non-educational problems in the opinion dynamics literature, including deterministic models with zero uncertainty, where $x_i(t,\theta)$ is a Dirac delta function \cite{deffuant_mixing_2000, degroot_reaching_1974, sood_voter_2005}, and probabilistic models like the one in this paper \cite{low_discerning_2022,low_vacillating_2022,bu_discerning_2023,jadbabaie_non-bayesian_2012,fang_opinion_2020,fang_social_2019}. 
In this section, and in Section \ref{sec:allies}, every node in the trust network is a student. 
We postpone introducing the teacher into the trust network until Section \ref{sec:partisan_student}.

\subsection{Observing assessment outcomes}
\label{subsec:observing_outcomes}

In the first step of the update rule, the $i$-th student applies Bayes's theorem to the score $0\leq S_i(t) \leq 1$ they receive for the $t$-th assessment task to update their prior belief $x_i(t,\theta)$ about the proficiency of the AI tool. 
Specifically, we write
\begin{equation}\label{eq:updatefirsthalf_ai}
    x'_i(t+1/2, \theta) = \frac{P[S_i(t) | \theta]}{P[S_i(t)]}x_i (t, \theta),
\end{equation}
where $x'_i(t+1/2, \theta)$ is the intermediate, provisional posterior after step one of the update rule, $P[S_i(t) | \theta]$ is the Bayesian likelihood, and the normalizing denominator in Eq.\ \eqref{eq:updatefirsthalf_ai} is the Bayesian evidence. 

Let $b_i(t)$ be a Boolean variable, which is true if the $i$-th student seeks AI assistance and false otherwise.
We call students who use AI (i.e.\ $b_i(t) = {\rm true}$) AI-users, and students who do not use AI (i.e.\ $b_i(t) = {\rm false}$) AI-avoiders. 
We assume for simplicity that the expectation value of $S_i(t)$ maps linearly to proficiency, such that zero and unit proficiencies map to zero and unit scores respectively on average, and the interpolation to intermediate values is linear. 
Upon including the unavoidable uncertainty generated by extraneous random factors, which affects the performance of the AI tool and the students, we assume that the scores are drawn from the conditional Gaussian distribution
\begin{align} \label{eq:Gaussian}
    S_i(t) \sim 
    \begin{cases}
        {\cal N}_{\rm T}(\theta_{\rm AI}, \sigma_{\rm AI}) &\text{ if } b_i(t) = {\rm true} \\ 
        {\cal N}_{\rm T}(\theta_{0,i},\sigma_{0}) &\text{ if } b_i(t) = {\rm false.} 
    \end{cases}
\end{align}
In Eq.\ \eqref{eq:Gaussian}, ${\cal N}_{\rm T}(y,z)$ denotes a Gaussian distribution with mean $y$ and standard deviation $z$ truncated to the domain $[0,1]$, $\sigma_{\rm AI}$ quantifies the spread in scores received by the AI on a given assessment, if the assessment task is repeated hypothetically many times in a short period\footnote{
    LLMs evolve with time as they learn from users' queries and responses \cite{kaufmann_survey_2023} as well as their own response \cite{huang_large_2022}. 
    Proficiency evolution is neglected in this paper as a first pass. 
    It happens during the fine-tuning step in LLM training and typically occurs a few months after user feedback \cite{brown_language_2020,openai_gpt-4_2024}.
    }, 
$\theta_{0,i}$ denotes the intrinsic proficiency of the $i$-th student, and $\sigma_0$ quantifies the spread in scores received by an AI-avoider, if the assessment task is repeated hypothetically many times.
The dispersion $\sigma_0$ reflects the many random factors that affect a student's performance in a specific random realization of an assessment task, such as their psychological state on the day, the detailed match between the task and their preparation, and so on. We assume that $\sigma_0$ is equal for every student for the sake of simplicity.
Eq.\ \eqref{eq:Gaussian} describes the recipe for generating synthetic scores for the Monte Carlo multi-agent simulations presented in Sections \ref{sec:allies} and \ref{sec:partisan_student}.

The Bayesian likelihood follows directly from Eq.\ \eqref{eq:Gaussian}. 
If the student spurns AI assistance, their score offers no guidance regarding the proficiency of the AI tool, and the likelihood is unity. If the student seeks AI assistance, their score is determined entirely by the AI tool's proficiency in the simplest version of the model through the first line of Eq.\ \eqref{eq:Gaussian}.
Therefore we write 
\begin{align}\label{eq:likelihood}
    P[S_i(t) | \theta] = 
    \begin{cases}
        \exp\bigl\{-[S_i(t) - \theta]^2/(2 \sigma_{\rm AI}^2)\bigr\}H[S_i(t)] H[1-S_i(t)] &\text{if } b_i(t) = {\rm true} \\
         1 &\text{if } b_i(t) = {\rm false,}
    \end{cases}
\end{align}
where $H(\dots)$ denotes the Heaviside step function.

Several aspects of Eqs.\ \eqref{eq:updatefirsthalf_ai}--\eqref{eq:likelihood} are more general than they need to be for the simplest version of the model examined in this paper. 
This is done deliberately, both to highlight the idealizations and approximations made here, and also to prepare the ground for generalizing the model in forthcoming papers. 
First, $b_i(t)$ is constant in time and fixed at the outset in this paper. 
Clearly, more complicated opinion dynamics may occur, if $b_i(t)$ is liberated to evolve exogenously (in response to external influences, e.g.\ social norms about the ethics of using AI) or endogenously (in response to the success or failure experienced by a student when using AI or not). 
Second, the students are only engaged in inferring $\theta_{\rm AI}$ in this paper, so $\theta_{0,i}$ in Eq.\ \eqref{eq:Gaussian} is not used in practice; it does not enter the inference chain connecting Eqs.\ \eqref{eq:updatefirsthalf_ai} and \eqref{eq:likelihood}. 
A refined model may incorporate $\theta_{0,i}$ in several meaningful ways: 
(i) students delegate the assessment task partially to the AI tool, so that $S_i(t)$ reflects $\theta_{\rm AI}(t)$ and $\theta_{0,i}(t)$ in some plausible combination\footnote{
    Realistically, a student may disregard partially the AI tool's solution, if the solution is unsatisfactory or much better than what the student can do on their own (to avoid being sanctioned by their teacher, if AI assistance is forbidden). 
    };
(ii) students adjust their strategy $b_i(t)$ from one time step to the next, in response to their beliefs about the relative proficiencies of the AI tool and themselves, i.e.\ by comparing where $\theta_{0,i}(t)$ lies within $x_i(t,\theta)$\footnote{
    Students may also adjust $b_i(t)$ exogenously, as social and educational norms evolve for or against AI assistance.
    };
(iii) students infer a PDF for their own proficiency if they are uncertain about it; 
and (iv) $\theta_{0,i}(t)$ evolves, as the use of the AI tool enhances or harms every student's intrinsic proficiency.
Third, in this paper, it is irrelevant whether $b_i(t)$ and $\theta_{0,i}(t)$ are public or not. 
In the second step of the update rule, described in Section \ref{subsec:peer_pressure}, students share their PDFs openly with everybody in the network. 
How the $i$-th student learns from the $j$-th student's PDF depends on whether the trust relationship between $i$ and $j$ is positive or negative (or whether they are related at all). 
One can imagine refined models, where one or more of the PDFs $x_i'(t+1/2,\theta)$ and $x_j'(t+1/2,\theta)$ and the variables $b_i(t)$, $b_j(t)$, $\theta_{0,i}(t)$, and $\theta_{0,j}(t)$ are hidden, and a student infers others' opinions based on partially revealed information \cite{jackson_social_2008}. 
This type of refinement lies outside the scope of the present paper, whose focus is on the role of trust networks {\em per se} instead of the partial transparency they may foster.

\subsection{Peer pressure in a trust network}
\label{subsec:peer_pressure}

After AI-users observe their score and infer the provisional posterior $x_i'(t+1/2,\theta)$, all students share their belief PDFs and arrive at a revised posterior $x_i(t+1,\theta)$ at the end of the time step. 
The revision is done according to trust relationships: allies trust each other, and their PDFs tend to converge, while opponents mistrust each other, and their PDFs tend to diverge.
The trust relationship between students $i$ and $j$ is described by the adjacency matrix, whose entries are given by
\begin{equation}
    A_{ij} = 
    \begin{cases}
        + 1 , \quad & \text{agents $i$ and $j$ are allies} \\
        0 , \quad & \text{agents $i$ and $j$ are disconnected} \\
        - 1 , \quad & \text{agents $i$ and $j$ are opponents.} \\
    \end{cases}
\end{equation}
The posterior belief PDF of agent $i$ updates according to
\begin{equation}\label{eq:updatesecondhalf}
    x_i(t+1, \theta) \propto \max\left[0, x_i '(t+1/2, \theta) + \mu \Delta x_i '(t + 1/2, \theta)\right],
\end{equation}
where 
\begin{equation}\label{eq:xiprimed}
    \Delta x_i '(t + 1/2, \theta) = a_i^{-1} \sum_{i\neq j} A_{ij}[x_j '(t+1/2, \theta) - x_i '(t+1/2, \theta)]
\end{equation}
denotes the average difference in belief (at each $\theta$) between agent $i$ and all its neighbors, and we have $a_i = \sum_{j \neq i} |A_{ij} |$.
The learning rate is restricted to the range $0.0 < \mu \leq 0.5$ to prevent overshooting, as discussed in Section 2.3 in \refcite{low_discerning_2022}.
For this paper, we take $\mu = 0.25$ without loss of generality.
The proportionality constant in Eq.\ \eqref{eq:updatesecondhalf} is set by normalizing the PDF.

\subsection{Relation to existing models}
\label{subsec:relation_to_existing_models}

To the best of our knowledge, this paper is the first to analyze students' opinions about the perceived proficiency of AI tools in the context of a trust network. 
Opinion dynamics models have also been constructed, in which AI tools participate endogenously as agents \cite{betz_natural-language_2022}. 
Their focus is to simulate collective deliberation by submitting new intellectual contributions in natural language (English), such as deductive arguments and empirical claims.
It is found that agents who actively submit new contributions dominate the evolution of the conversation. 
In this paper, in contrast, AI tools feature passively as entities about which humans form opinions.

Previous opinion dynamics models that focus on education treat teachers as leaders, i.e.\ as influential nodes with high degree \cite{koeslag-kreunen_leadership_2018,li_improved_2013}. 
It is found that students reach consensus and agree with their teachers, when all teachers hold the same belief.
The perception of higher teaching ability can drive students towards a consensus more effectively \cite{zhou_effects_2018}.
The findings about teachers apply to AI-related issues in principle too but are not derived specifically in an AI context. 

Many models in the opinion dynamics literature consider deterministic beliefs, where each agent holds a single belief at each time step with 100\% certainty, i.e.\ $x_i(t, \theta_1) = 1$ and $x_i(t, \theta \neq \theta_1) = 0$; see Refs.\ \cite{sood_voter_2005,degroot_reaching_1974,deffuant_mixing_2000} for examples.
Several deterministic models adopt deterministic analogs of the probabilistic linear update rule in Eq.\ \eqref{eq:updatesecondhalf},  for example the Deffuant-Weisbuch model \cite{deffuant_mixing_2000}.
In this paper, we generalize the beliefs of agent $i$ to be multi-valued and therefore uncertain. 
Each student may hold different levels of confidence about two or more distinct values of $\theta_{\rm AI}$, e.g.\ $x_i(t,\theta_1)=0.4$ and $\sum_{\theta_2} x_i(t,\theta_2 \neq \theta_1) = 0.6$. 
Some other models in the literature likewise represent an agent's opinion as a  PDF of nonzero width, including Refs.\ \cite{fang_opinion_2020,jadbabaie_non-bayesian_2012,low_discerning_2022,low_vacillating_2022,bu_discerning_2023}.
Opponents in a network have been studied with multi-valued belief PDFs \cite{low_discerning_2022,low_vacillating_2022,bu_discerning_2023} as well as with deterministic models, which focus on when and why opinions diverge \cite{vaz_martins_mass_2010,shi_evolution_2016,chen_opinion_2019,he_discrete-time_2021}.

\section{Perceived proficiency in the long term}
\label{sec:allies}
We first study the long-term evolution of students' opinions about an AI tool's proficiency in different network settings.
To do so, we perform Monte Carlo simulations based on iterating Eqs.\ \eqref{eq:updatefirsthalf_ai}--\eqref{eq:xiprimed}. 
In line with human psychology \cite{tee_quantized_2019}, the continuous variable $\theta$ is discretized into 21 regularly-spaced values, $\theta \in \{0.00, 0.05, \dots , 1.00\}$.
In the rest of the paper, unless stated otherwise, we take $\theta_{\rm AI} = 0.8$ and $\sigma_{\rm AI} = 0.2$\footnote{
    In this paper, we do not need values for $\theta_{0,i}$ and $\sigma_0$ in Eq.\ \eqref{eq:Gaussian}, as explained in Section \ref{subsec:observing_outcomes}.
    The parameters $\theta_{0, i}$ and $\sigma_0$ are preserved in Eq.\ \eqref{eq:Gaussian} as a formality to flag in advance a framework for generalizing the model in a forthcoming paper.
}. 
When modeling LLM-based AI tools, for example, $\sigma_{\rm AI} \neq 0$ can be related to the temperature hyperparameter of the LLM, which inflates stochastically the probability of outputting low-probability tokens (words) to avoid generating repetitive and unimaginative text by introducing some randomness and diversity in the output \cite{openai_gpt-4_2024,fan_hierarchical_2018,holtzman_curious_2020,radford_language_2019}.
In addition, an LLM responds somewhat stochastically to the wording of a prompt \cite{yang_large_2024}, e.g.\ from users with different sociodemographic backgrounds \cite{chen_conversational_2024}. 

In Section \ref{subsec: correct_allies_only}, we simulate allies-only networks and find that all students correctly infer $\theta_{\rm AI}$, but AI-users and AI-avoiders reach their conclusions at different speeds.
In Section \ref{subsec: incorrect_opponent_only}, we simulate opponents-only networks and find that every student tends to a steady, unimodal, asymptotic belief PDF, but some students infer $\theta_{\rm AI}$ incorrectly.
In Section \ref{subsec: turbulent_nonconvergence}, we simulate networks with mixed trust relationships and find that some or all of the students exhibit turbulent noncovergence, depending on their location in the network, as discovered previously in noneducational applications \cite{low_discerning_2022,low_vacillating_2022,bu_discerning_2023,fang_social_2019,lalitha_social_2018}. 

\subsection{Correctly inferring $\theta_{\rm AI}$ in allies-only networks}
\label{subsec: correct_allies_only}
Let us start by considering a complete allies-only network with $n = 10$, i.e.\ a relatively small class or study group, where mutual trust is high due to friendship bonds or esprit de corps.
All students are connected the same way within this network. 
We examine all possible configurations of AI-users and AI-avoiders: for $1 \leq k \leq n-1$, we randomly choose $k$ students to be AI-users, and $n - k$ students to be AI-avoiders\footnote{
    We do not consider the case $k =0$, where the network transmits no information about the AI's proficiency, because our goal is to infer $\theta_{\rm AI}$. 
    We do not show results for $k = n$, where all students are AI-users, because our goal is to study the difference in long-term behavior between AI-users and AI-avoiders. 
}.
We run 100 simulations for each configuration with randomized prior beliefs and score sequences (with $T=10^4$). 
We note that the long-term outcomes of iterating the update rules given by Eqs.\ \eqref{eq:updatefirsthalf_ai}--\eqref{eq:xiprimed} do not depend strongly on the priors and score sequence, as demonstrated systematically in Appendix C in Ref.\ \cite{bu_discerning_2023}. 
When classifying the long-term behavior of the system, we distinguish between \textit{consensus} and \textit{asymptotic learning}. 
We say that students $i$ and $j$ reach consensus if they agree with each other, viz.\
\begin{equation}\label{eq:consensus}
    \text{max}_{\theta} |x_i(t,\theta) -x_j(t,\theta) | < \epsilon \text{ max}_{\theta} \left[ x_i(t, \theta), x_j(t, \theta) \right], 
\end{equation}
where $\epsilon = 0.01$ is a user-selected tolerance. 
Consensus within a group of students is reached, if Eq.\ \eqref{eq:consensus} is satisfied for all pairs of students within that group.
We say that student $i$ achieves asymptotic learning, if their PDF remains unchanged within a user-selected tolerance for a specified period of time, $\tau_{\rm max}$, viz.\
\begin{equation} \label{eq:asymlearncondition}
    \max_{\theta} |x_i(t+\tau, \theta) - x_i(t, \theta)| < \epsilon \max_{\theta} |x_i(t, \theta)| 
\end{equation}
for $ 1 \leq \tau \leq \tau_{\rm max}$.
In this paper, we take $\tau_{\rm max} = 99$ typically, copying the arbitrary choice made in Ref.\ \cite{low_discerning_2022,low_vacillating_2022,bu_discerning_2023}.
The asymptotic learning time, $t_{\rm a}$, is the minimum $t_{\rm a}$ value for which Eq.\ \eqref{eq:asymlearncondition} is satisfied for $t \geq t_{\rm a}$.
The automaton iterates, until the maximum run-time $T$ is reached.  
We emphasize that consensus is not the same as asymptotic learning, nor does one imply the other. 
Students' beliefs may vacillate, while maintaining consensus; conversely, students may achieve stationary beliefs while disagreeing with each other.

\begin{figure}[h]
    \centering
    \begin{subfigure}{0.6\textwidth}
        \includegraphics[width=\linewidth]{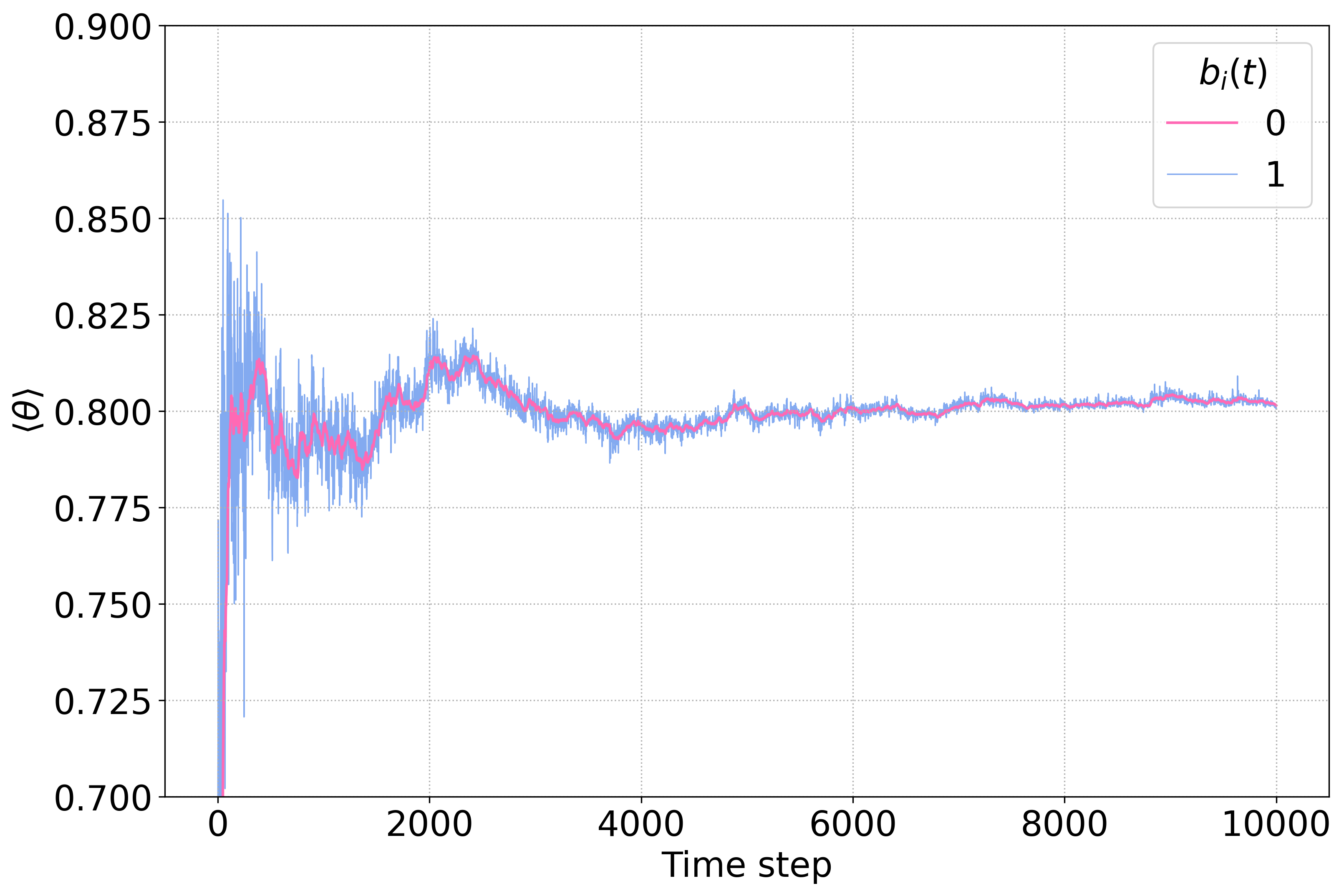}
    \end{subfigure}
    \\
    \begin{subfigure}{0.5\textwidth}
        \includegraphics[width=\linewidth]{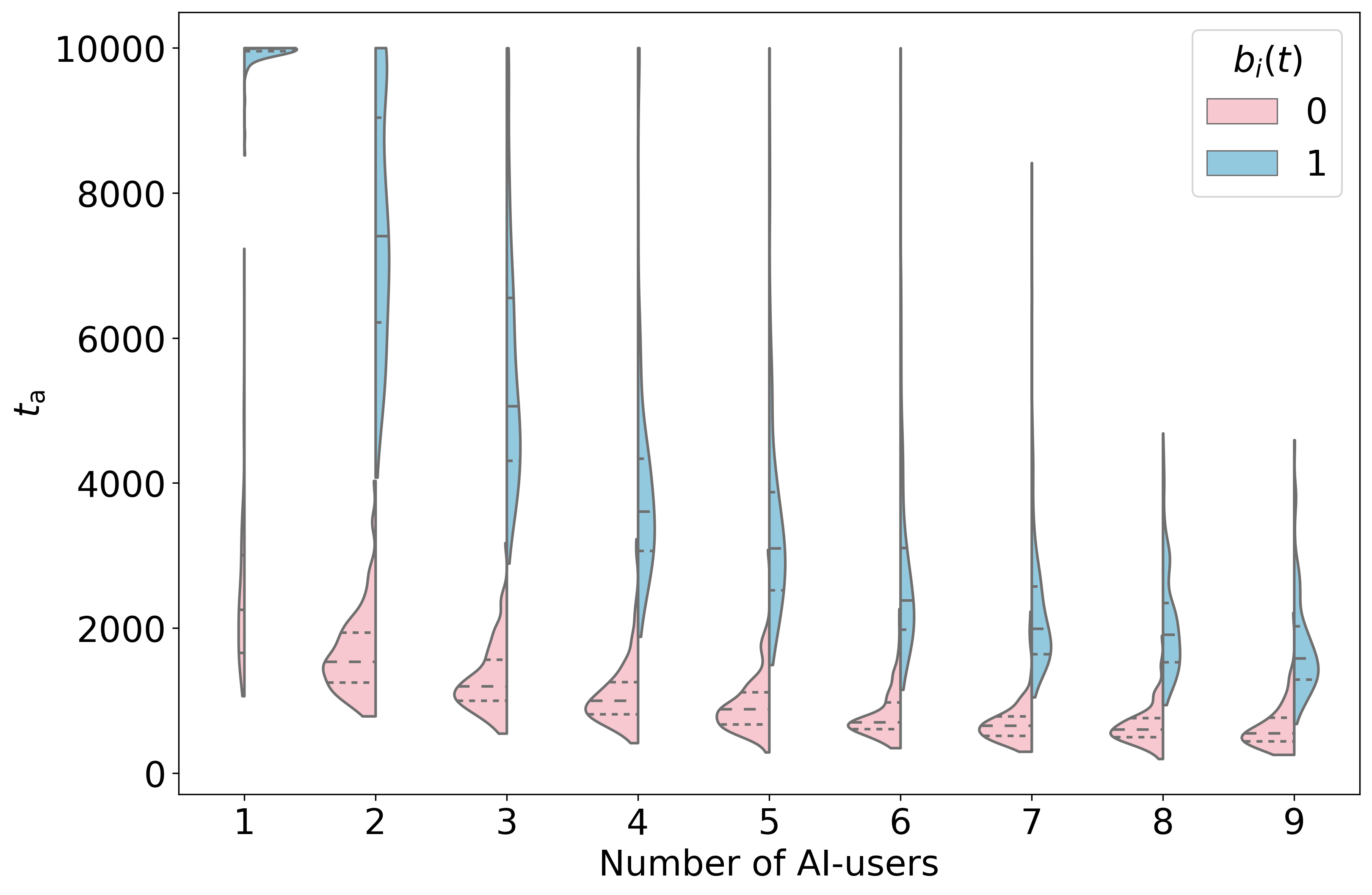}
    \end{subfigure}%
    \hspace*{\fill}  
    \begin{subfigure}{0.5\textwidth}
        \includegraphics[width=\linewidth]{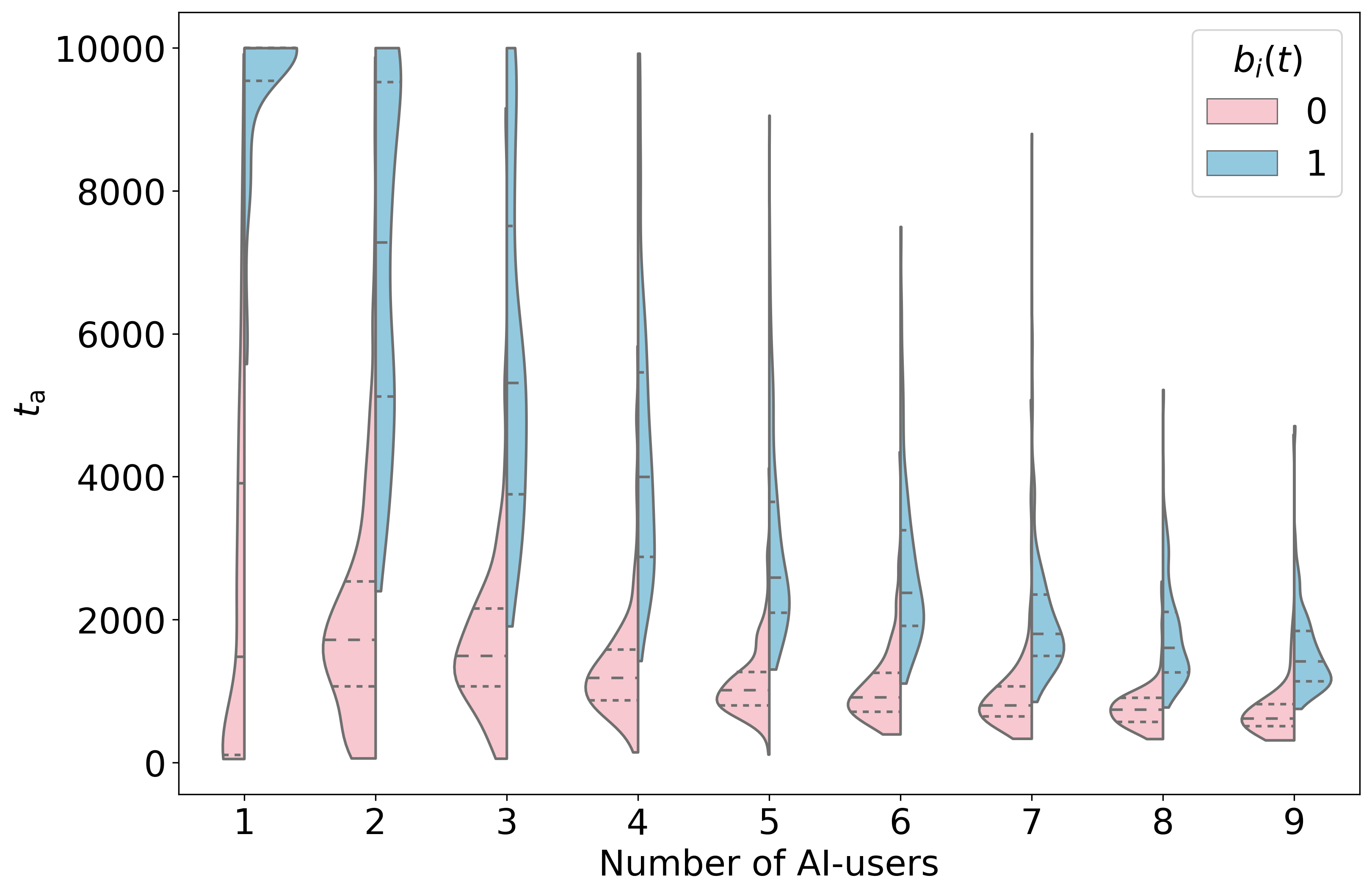}
    \end{subfigure}
    \caption{
        Correctly inferring $\theta_{\rm AI}$ in an allies-only network with $n=10$. 
        (Top.) 
            Mean belief $\langle \theta \rangle$ versus time $t$ for a representative AI-user (blue curve) and AI-avoider (pink curve). 
            Both curves asymptote to the correct value of $\theta_{\rm AI}$.
        (Bottom left.) 
            Semi-violin plot of the distribution of asymptotic learning time $t_{\rm a}$ as a function of the number of AI-users in a complete network.
            Histograms of $t_{\rm a}$ are separated into AI-users ($b_i(t) = 1$, blue) and AI-avoiders ($b_i(t) = 0$, pink).
            The horizontal width (left or right) of the semi-violin plot represents the number of students in a $t_{\rm a}$ bin (vertical axis), normalized to give the same area for each half of the violin.
            Dotted horizontal lines indicate the quartiles of $t_{\rm a}$.       
        (Bottom right.) 
            As for the bottom left panel, but for a Barab\'{a}si-Albert network. 
            In the bottom two panels, we run 100 simulations with randomized priors and AI-user locations for $1\leq k \leq 9$, amounting to 900 simulations in total.
            AI-avoiders achieve asymptotic learning faster than AI-users. 
        The difference in $t_{\rm a}$ between AI-users and AI-avoiders decreases, as the number of AI-users increases. 
        }\label{fig:diff_t_A}
\end{figure}

The top panel of Fig. \ref{fig:diff_t_A} shows the mean belief $\langle \theta \rangle_{i}(t) = \int_0^1 d \theta' \theta' x_i(t, \theta')$, abbreviated to $\langle \theta \rangle$, for a representative AI-user (blue curve) and a representative AI-avoider (pink curve) in a simulation with $k = 1$. 
Both the AI-user and the AI-avoider learn asymptotically the injected value $\theta_{\rm AI}=0.8$. 
They do so at roughly the same time, with $| \langle \theta \rangle - \theta_{\rm AI} | < 0.01$ for $t \gtrsim 3\times 10^3$ for both agents, although it turns out that AI-avoiders learn faster than AI-users on average, when one analyzes $t_{\rm a}$ more closely (see below). 
Furthermore, the root-mean-square fluctuations are greater for the AI-user ($\approx 5 \times 10^{-3}$) than for the AI-avoider ($\approx 4 \times 10^{-3}$).  
The simulations reveal that AI-avoiders reach consensus among themselves quickly ($t \gtrsim 20$), and remain in consensus until $t = T$. 
In complete networks, the connections between all students are identical, which fosters consensus. 
Furthermore, allies with the same belief do not influence each other through Eq.\ \eqref{eq:updatesecondhalf}.
At $t = T$ for all simulations, all students infer correctly the proficiency of the AI tool.
We find $x_i(t=T, \theta = \theta_{\rm AI}) > 0.95$ and $x_i(t=T, |\theta - \theta_{\rm AI}| > 0.05) = 0$.
The belief PDFs are approximately stationary for $t\gtrsim 3\times 10^3$, exhibiting fluctuations below the threshold defined by Eq.\ \eqref{eq:asymlearncondition}.

Counterintuitively, we find that AI-avoiders infer $\theta_{\rm AI}$ correctly and achieve asymptotic learning faster than AI-users, despite lacking direct access to the AI scores.
The bottom left panel of Fig.\ \ref{fig:diff_t_A} shows the distribution of $t_{\rm a}$ for AI-users ($b_i(t) = 1$) and AI-avoiders ($b_i(t) = 0$) in a complete network.
The panel displays nine semi-violin plots, corresponding to $1\leq k \leq 9$, whose horizontal extent (left or right) is proportional to the count of $t_{\rm a}$ values (out of 900) in the associated $t_{\rm a}$ bin plotted on the vertical axis. 
For example, for $n=1$, the histogram peaks at $t_{\rm a} \approx 1.0 \times 10^4$ for AI-users (blue histogram) and $t_{\rm a} \approx 2.3 \times 10^3$ for AI-avoiders (pink histogram). 
Two trends are clear. The blue histograms peak at systematically higher $t_{\rm a}$ than the pink histograms for all $k$, and the modal $t_{\rm a}$ value decreases monotonically with $k$.
AI-users are exposed directly to the Gaussian fluctuations in $S_i(t)$, whereas AI-avoiders learn about $S_i(t)$ indirectly through the network average described by the sum in Eq.\ \eqref{eq:xiprimed}, whose fluctuations in percentage terms are $(n-1)^{1/2}$ times smaller by the central limit theorem.
As the number of AI-users increases,  $t_{\rm a}$ for AI-users remains larger on average than $t_{\rm a}$ for AI-avoiders, because AI-users as a group have direct access to more scores per time step. 
On the other hand, $t_{\rm a}$ for AI-avoiders depends weakly on $k$, because the network average in Eq.\ \eqref{eq:updatesecondhalf} involves $n-1$ terms independent of $k$.
In each semi-violin plot, the two semi-histograms overlap, because AI-users have strictly larger $t_{\rm a}$ than AI-avoiders in the same simulation but may have equal or smaller $t_{\rm a}$ in a different simulation. 
The above effects are analogous loosely to the celebrated social science phenomenon, that one can predict the result of an election more accurately by polling people's predictions about other people's voting intentions rather than their own \cite{rothschild_forecasting_2011}.

We repeat the experiment in Fig. \ref{fig:diff_t_A} in a Barab\'{a}si-Albert network with $n = 10$, attachment parameter $m = 3$, and all possible numbers of AI-users $1 \leq k \leq n-1$.
The aim is to check if the results in Fig. \ref{fig:diff_t_A} are independent of the network topology.
Unlike in complete networks, the connections differ for each node, so we randomize the location of AI-users in each simulation. 
We find that AI-avoiders no longer reach consensus among themselves, as their connections differ.
However, we still observe that $t_{\rm a}$ for AI-users typically exceeds $t_{\rm a}$ for AI-avoiders, as seen in the semi-violin plots in the bottom right panel of Fig.\ \ref{fig:diff_t_A}.
For example, for $k=3$, the blue histogram (AI-users) peaks at $t_{\rm a} \approx 5.3 \times 10^{3}$, while the pink histogram (AI-avoiders) peaks at $t_{\rm a} \approx 1.5 \times 10^{3}$.
Moreover, $t_{\rm a}$ for AI-users decreases, as $k$ increases, replicating the trend in complete networks.
Students in a Barab\'{a}si-Albert network typically achieve asymptotic learning slower than in a complete network, whether they are AI-users or AI-avoiders, because information is transmitted slower, when there are fewer connections overall.

\subsection{Incorrectly inferring the AI's proficiency in opponents-only networks}
\label{subsec: incorrect_opponent_only}
\begin{figure}[h]
    \centering
    \begin{subfigure}{0.5\textwidth}
        \centering
        \includegraphics[width=\linewidth]{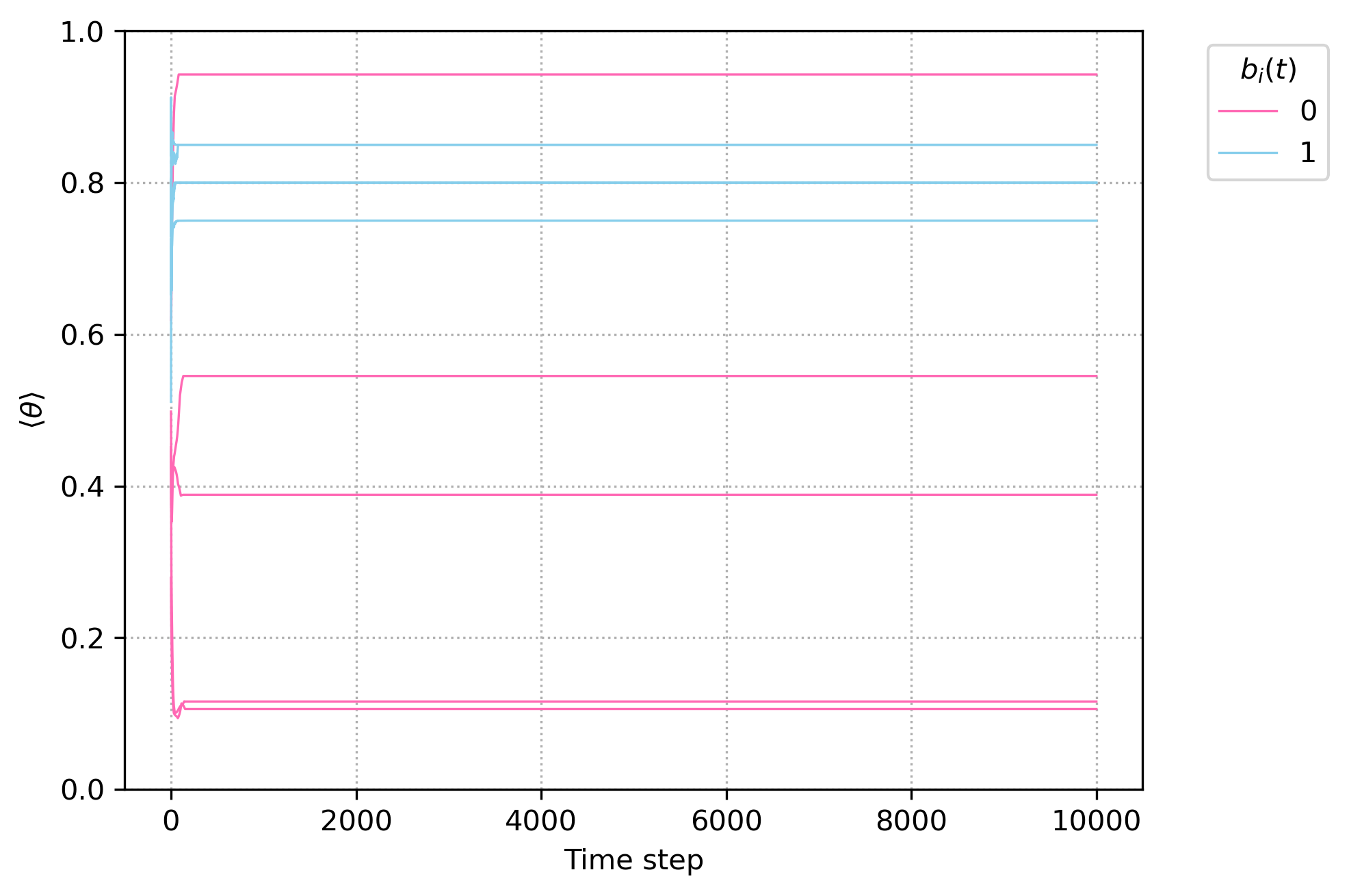}
    \end{subfigure}%
    \hspace*{\fill}  
    \begin{subfigure}{0.5\textwidth}
        \centering
        \includegraphics[width=\linewidth]{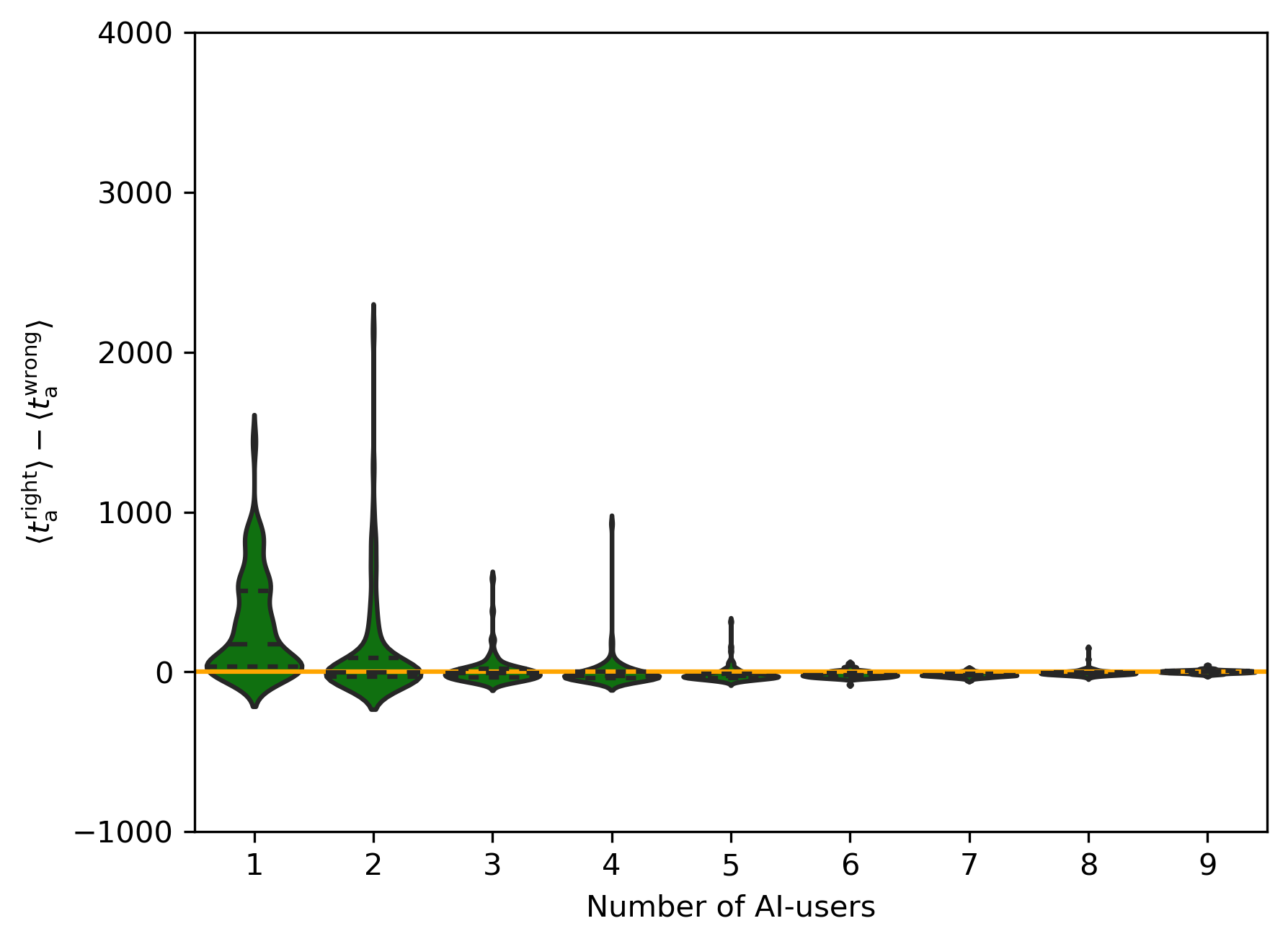}
    \end{subfigure} \\
    \begin{subfigure}{0.5\textwidth}
        \centering
        \includegraphics[width=\linewidth]{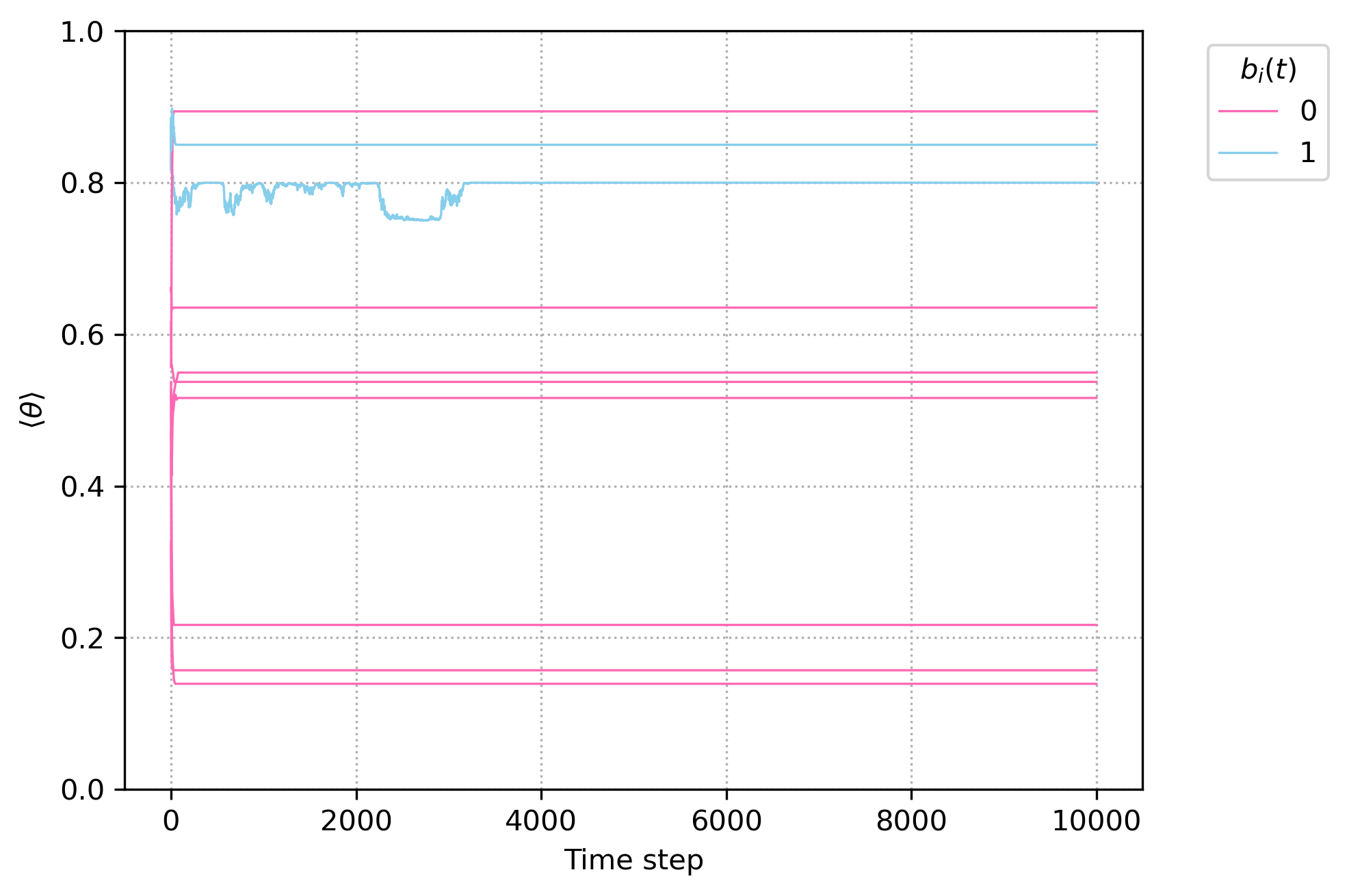}
    \end{subfigure}%
    \hspace*{\fill}  
    \begin{subfigure}{0.5\textwidth}
        \centering
        \includegraphics[width=\linewidth]{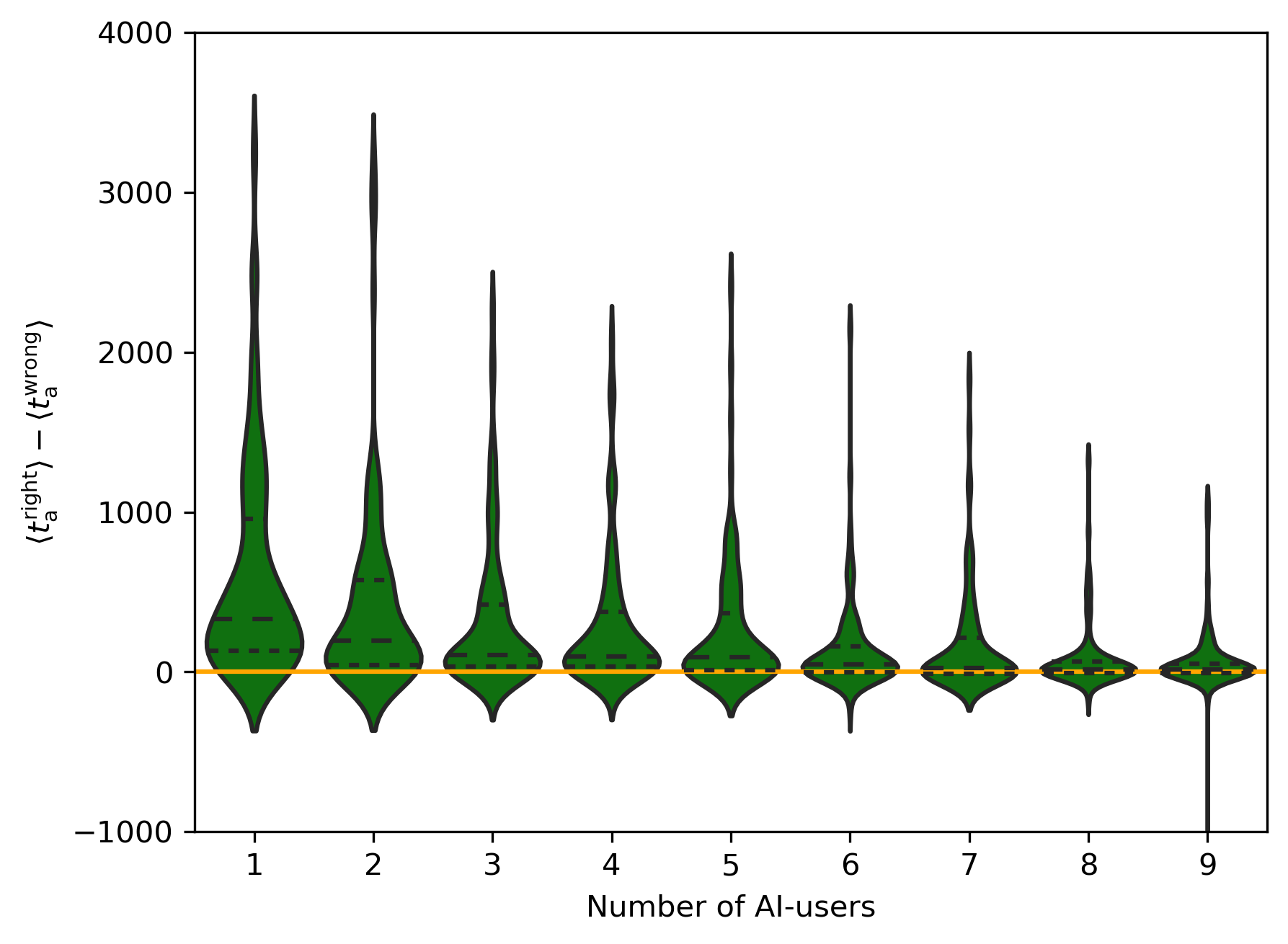}
    \end{subfigure}%
    \caption{
        Incorrectly inferring $\theta_{\rm AI}$ in an opponents-only network with $n=10$.
    (Top left.) 
        Mean belief $\langle \theta \rangle$ versus time $t$ for AI-users (blue curve) and AI-avoiders (pink curve). 
        Two of the AI-users and none of the AI-avoiders  infer $\theta_{\rm AI}$ correctly.
    (Top right.)
        Violin plot of the distribution of the difference in asymptotic learning time $\langle t_{\rm a}^{\rm right} \rangle - \langle t_{\rm a}^{\rm wrong} \rangle$ between students who do and do not correctly infer $\theta_{\rm AI}$, averaged over the $n$ students in each simulation, as a function of the number of AI-users.
        The horizontal orange line indicates no difference in asymptotic learning time, i.e.\ $ \langle t_{\rm a}^{\rm right} \rangle = \langle t_{\rm a}^{\rm wrong} \rangle $.
        (Bottom left and bottom right.)
        Same as the top two panels, but for a Barab\'{a}si-Albert network with $n = 10$ and $m = 3$.
    }
    \label{fig:opponent}
\end{figure}

We now consider opponents-only networks with $n =10$.
One realistic example is a small class of postgraduate coursework students competing in an adversarial climate for a limited resource, e.g.\ to win a scholarship to fund further postgraduate study. 
We first analyze complete networks, with $1 \leq k \leq 9$ AI-users and $n - k$ AI-avoiders, and test whether the findings in Section \ref{subsec: correct_allies_only} persist. 
We run 100 simulations for each $k$, totaling 900 simulations. 
The results are summarized in Fig.\ \ref{fig:opponent}.

In the long run, all students in a complete, opponents-only network achieve asymptotic learning, but not all students infer the correct $\theta_{\rm AI}$. 
The top left panel of Fig.\ \ref{fig:opponent} graphs $\langle \theta\rangle$ as a function of time for all 10 students in a typical simulation with $k = 5$. 
Two of the five AI-users and none of the five AI-avoiders infer the correct $\theta_{\rm AI}$, and every student achieves asymptotic learning at $t_{\rm a} = 126$. 
After asymptotic learning is achieved, the belief PDFs of AI-users are single-valued, i.e. $ x_i(t, \theta)$ is only non-zero at a particular value of $\theta$ driven by $S_i(t)$. 
Some AI-users infer $\theta_{\rm AI}$ incorrectly due to antagonistic interactions between opposed AI-users.
Similar behavior is observed in Refs.\ \cite{low_discerning_2022,bu_discerning_2023}.
AI-avoiders do not have single-valued belief PDFs, as their beliefs evolve though peer pressure to minimize overlap with their opponents. 
The beliefs of AI-avoiders usually do not cluster around $\theta_{\rm AI}$, as  they are repelled by opposed AI-users at $\theta_{\rm AI}$.
Across all simulations, a total of 1479 (out of 9000) students correctly infer $\theta_{\rm AI}$, of whom only five are AI-avoiders from five different simulations.
The latter five simulations have $1\leq k \leq 2$ AI-users, who fail to infer $\theta_{\rm AI}$ correctly and therefore leave the belief $\theta=\theta_{\rm AI}$ open for AI-avoiders to ``occupy''. 

The above behavior has two important educational implications.
(i) Students in a low-trust environment are likely to form erroneous perceptions about the intrinsic proficiency of an AI tool, including whether the tool is more or less proficient than they are themselves when unassisted, thereby harming their ability to make sound, grade-maximizing judgments about its use.
(ii) It helps a little, but not a lot, for students to try and ``discover the truth for themselves'' by using the AI tool, because peer pressure from the network confounds the signal $S_i(t)$. 
Counterintuitively, property (ii) is somewhat egalitarian: it does not matter much in the end whether a student can access the AI tool or not (e.g.\ due to socioeconomic factors), because AI-users and AI-avoiders both struggle to assess the tool's proficiency correctly. However, AI-users are still privileged over AI-avoiders to some extent, and besides the entire situation (wrong perceptions are widespread) is unsatisfactory, as one expects when trust is low.

Ref.\ \cite{low_discerning_2022} finds another property in opponents-only networks: often the wrong conclusion is reached first.
That is, $t_{\rm a}$ for students who reach the wrong conclusion is less than $t_{\rm a}$ for students who reach the right conclusion ($t_{\rm a}^{\rm wrong} < t_{\rm a}^{\rm right}$). 
In a noneducational context, this property depends on network topology and only holds for sparse networks \cite{bu_discerning_2023}. 
Does it hold also for the education-inspired model in this paper?
Let us test the issue as follows.
For each complete network simulated in the top panels in Fig.\ \ref{fig:opponent}, we calculate $\langle t_{\rm A,right} \rangle$ and $\langle t_{\rm A,wrong} \rangle$ in every simulation. 
Here, $\langle \ldots \rangle$ denotes an average over all the students who reach the right and wrong conclusions respectively in a particular simulation.
The top right panel of Fig.\ \ref{fig:opponent} shows the distribution for 900 simulations of $\langle t_{\rm a}^{\rm right} \rangle - \langle t_{\rm a}^{\rm wrong} \rangle$ as nine violin plots, classified according to the number of AI-users in the network\footnote{
    We do not subdivide the histogram to separately count AI-users and AI-avoiders, as per Fig.\ \ref{fig:diff_t_A}, as both AI-users and AI-avoiders can reach either the right or wrong conclusion.}.
The dashed, black, horizontal lines indicate the quartiles, and the orange horizontal line shows $\langle t_{\rm a}^{\rm right} \rangle = \langle t_{\rm a}^{\rm wrong} \rangle$. 
For $k = 1$, we observe a tendency for the wrong conclusion to be reached first; the orange line aligns roughly with the dashed line for the first quartile.
For $k > 1$, in contrast, the orange line roughly aligns with the dashed line for the median, i.e.\ we do not observe wrong conclusions being reached first. 
Additionally, we find that AI-users tend to have longer $t_{\rm a}^{\rm right}$.
For $k =1$, at the beginning of the simulation, the antagonistic interaction pushes the AI-users to minimize overlap with their opponents' beliefs and $S_i(t)$. 
Their beliefs exhibit turbulent nonconvergence during this time, i.e.\ the belief vacillates without settling, as found in media bias applications \cite{low_vacillating_2022}.
As $t$ increases, AI-avoiders achieve asymptotic learning with a belief PDF, which minimizes overlap and depends partly on the prior regardless of $\theta_{\rm AI}$.
After some time, they no longer exert peer pressure on AI-users via Eq.\ \eqref{eq:updatesecondhalf}; that is, an AI-user's belief evolves solely and stochastically in response to $S_i(t)$ towards $\theta_{\rm AI}$, resulting in longer $t_{\rm a}^{\rm right}$. 
The switch from turbulent nonconvergence to asymptotic learning is often triggered by $S_i(t)$.
For example, in one of the simulations, we discover an AI-user whose score sequence $S_i( 564 \leq t \leq 566 ) = 0.92, 1, 1$ pushes $\langle \theta \rangle$ from $\langle \theta \rangle = 0.76$ to $\langle \theta \rangle = 0.8$,  but the agent does not achieve asymptotic learning straight afterwards; their belief vacillates until $t = 2215$, and as a result of the Gaussian fluctuations in $S_i(t)$.
However, as $k$ increases, more students' beliefs evolve toward $\theta_{\rm AI}$, and the antagonistic interactions around $\theta_{\rm AI}$ predominate.
Any drift towards $\theta=\theta_{\rm AI}$ caused by $S_i(t)$ is counteracted by $ \Delta x_i'(t + 1/2, \theta)$, which restores $x_i(t,\theta)$ towards stationarity, leading to shorter $t_{\rm a}$ for AI-users.


We now repeat the complete network experiments for Barab\'{a}si-Albert networks with $n = 10$, $m = 3$, and $1 \leq k \leq n-1$.
The bottom two panels in Fig.\ \ref{fig:opponent} show the evolution of students' mean beliefs with respect to time (left panel) and the distribution of $\langle t_{\rm a}^{\rm right} \rangle - \langle t_{\rm a}^{\rm wrong} \rangle$ (right panel).
The plot formats are the same as in the top two panels.
The results broadly resemble the top two panels.
Only one AI-user infers $\theta_{\rm AI}$ correctly.
Interestingly, the same student exhibits turbulent nonconvergence for $t \lesssim 3.2 \times 10^3$ before settling on the truth. 
This happens because the AI-user and one of the opposing AI-avoiders both have $x_i(t, \theta = \theta_{\rm AI} -0.05) \neq 0$, and simultaneously we find $x_{\text{AI-user}}(t, \theta = \theta_{\rm AI} -0.05) > x_{\text{AI-avoider}}(t, \theta = \theta_{\rm AI} -0.05)$ and hence $\Delta x_{\text{AI-user}}'(t + 1/2, \theta= \theta_{\rm AI} -0.05) > 0$.
The student enters a stationary state triggered by the specific score sequence $S_{\text{AI-user}}(2801 \leq t \leq 2804) = 1,1,1, 0.93$.

\subsection{Intermittency in mixed networks}
\label{subsec: turbulent_nonconvergence}

Intermittency is a phenomenon which emerges in mixed networks featuring both positive and negative trust relationships \cite{low_discerning_2022,low_vacillating_2022,bu_discerning_2023}. 
It occurs when an agent experiences turbulent nonconvergence for many time steps, during which they vacillate between two or more modal beliefs, after which the agent switches into a stable state for many time steps, during which $x_i(t,\theta)$ remains stationary. 
The cycle then repeats. 
Opinion dynamics models of media bias have elucidated some of the conditions which foster intermittency, e.g.\ network structure \cite{low_discerning_2022,low_vacillating_2022,bu_discerning_2023} as interpreted through the social science theory of structural balance \cite{heider_attitudes_1946,cartwright_structural_1956}.

In this paper, we also observe intermittency under various circumstances. 
Indeed, we even observe it in some of the opponents-only networks simulated to create Fig.\ \ref{fig:opponent}.
For example, in the simulation with $k=9$ that yields $\langle t_{\rm a}^{\rm right} \rangle - \langle t_{\rm a}^{\rm wrong} \rangle < -10^3$ in the $k=9$ histogram in the bottom right panel of Fig.\ \ref{fig:opponent}, we discover an intermittent AI-user. 
This particular student's belief PDF evolves as follows: 
(i) stability after an initial transient, with $x_i(t,\theta=\theta_{\rm AI})=1$ for $255 \leq t \leq 8372$; 
(ii) turbulent nonconvergence, with $x_i(t,\theta = \theta_{\rm AI}) > 0$ and $x_i(t,\theta = \theta_{\rm AI} + 0.05) > 0$ for $8373 \leq t \leq 9120$, triggered by $S_i(8373)=0.98$ and $S_i(8374)=1.0$; 
and (iii) stability again, with $x_i(t,\theta=\theta_{\rm AI}+0.05)=1$ for $t\geq 9121$, triggered by $S_i(9113 \leq t \leq 9120) = 0.99, 1, 1, 0.62, 1, 1, 0.60, 0.70$. 
The intermittency is driven by the interplay between two opponents at $\theta=\theta_{\rm AI}$ and another two opponents at $\theta=\theta_{\rm AI}+0.05$.

\begin{figure}[h!]
    \centering
    \begin{subfigure}{0.5\textwidth}
        \centering
        \includegraphics[width=\linewidth]{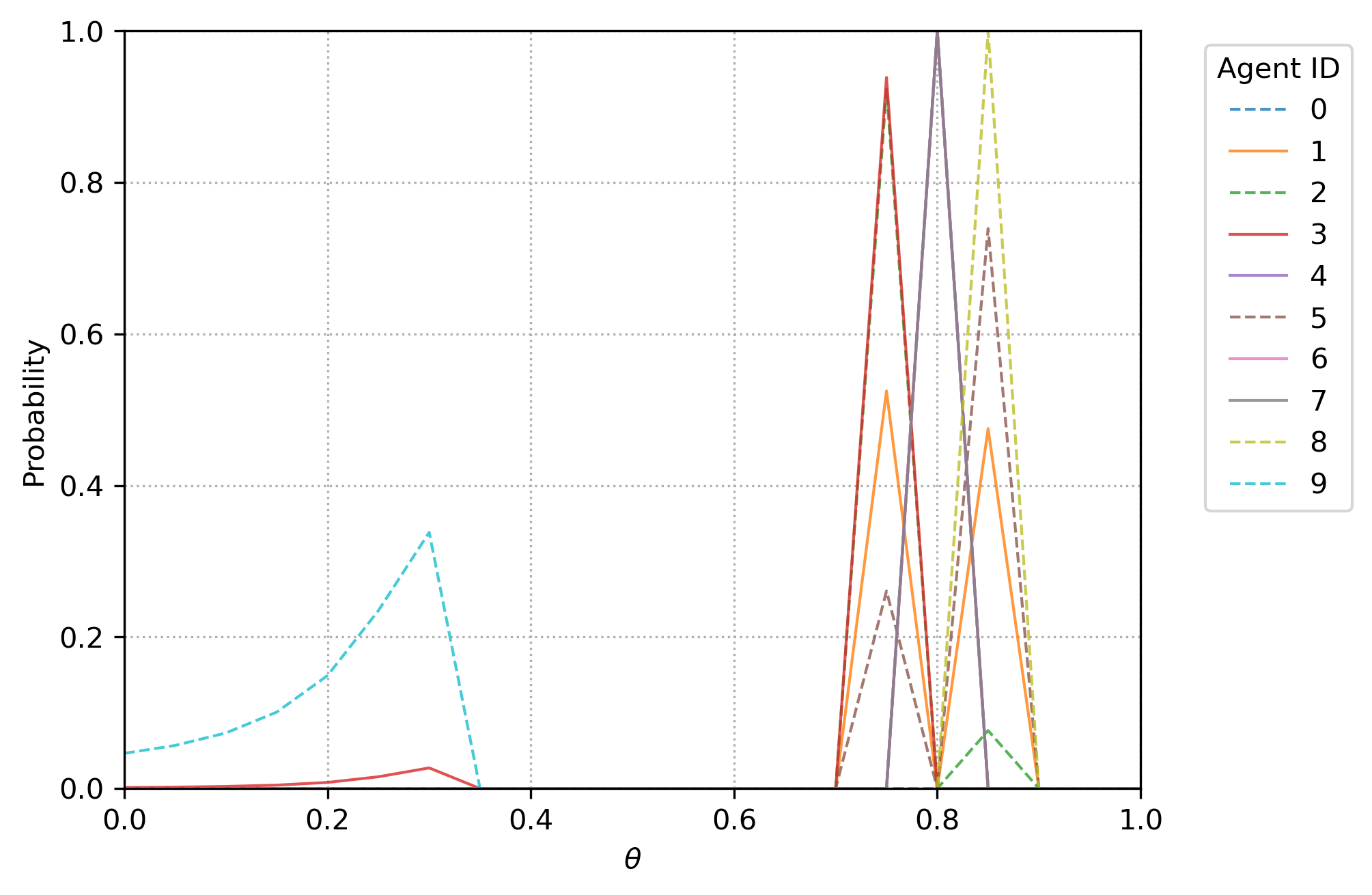}
    \end{subfigure}%
    \hspace*{\fill}  
    \begin{subfigure}{0.5\textwidth}
        \centering
        \includegraphics[width=\linewidth]{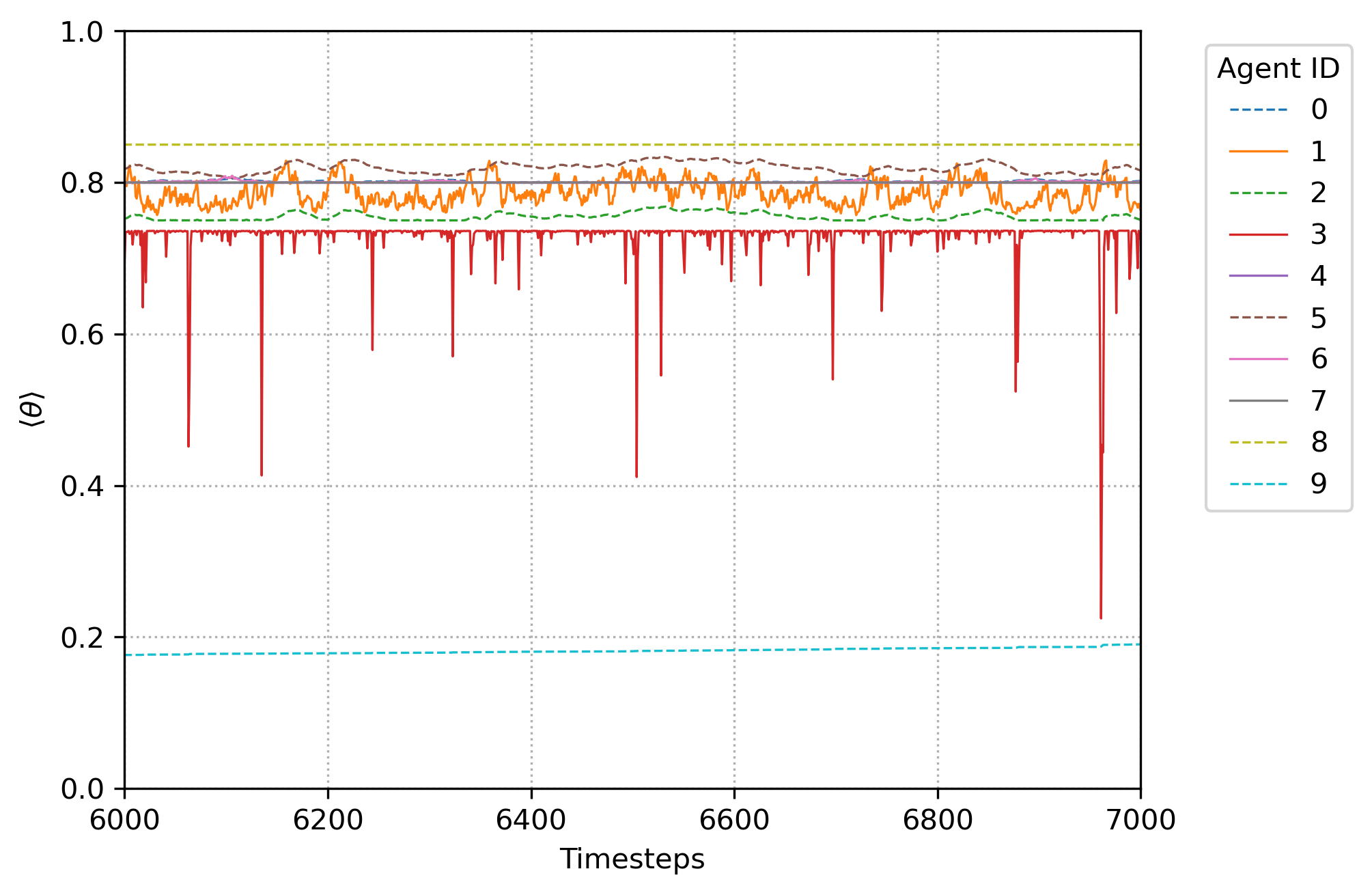}
    \end{subfigure} 
    \caption{
        Turbulent nonconvergence and intermittency in a mixed network, for a typical simulation of a complete network with $n = 10$ and $k=5$ AI-users.
        The relationship between each pair of students is generated independently and randomly, with equal probability for $A_{ij}=\pm 1$.
        AI-users are represented with solid curves, and AI-avoiders with dashed curves.
        (Left.)
            Snapshot of belief PDF at $t = 6\times 10^3$ for all 10 students. 
        (Right.)
            Mean $\langle \theta \rangle$ of the belief PDF for each student versus time for $6\times 10^3 \leq t \leq 7\times 10^3$. 
            Students 0, 1, 2, 3, 5, 6, 7 (four AI-users, three AI-avoiders) and 4, 8 (one AI-user, one AI-avoider) exhibit turbulent nonconvergence and asymptotic learning respectively.
        }
    \label{fig:mixed}
\end{figure}

To investigate intermittency in a mixed network, we choose one representative simulation of a complete network with $n=10$ and $k=5$ and examine the behavior of each student in detail. 
This example is illustrative; there are many alternatives of course, which are too numerous to study exhaustively.
The simulation features 20 pairs of allies and 25 pairs of opponents. 
The left panel of Fig.\ \ref{fig:mixed} displays a snapshot of all 10 belief PDFs at $t = 6\times 10^3$.
The right panel shows how $\langle \theta \rangle$ evolves for every student for $6 \times 10^3 \leq t \leq 7\times 10^3$. 
The dashed and solid curves represent AI-avoiders and AI-users respectively.
The following long-term behaviors are observed.

\begin{itemize}
    \item Student 4 (AI-user) and student 8 (AI-avoider) achieve asymptotic learning at $t_{\rm a} = 2965$ and $t_{\rm a} = 395$ and infer $\theta_{\rm AI}$ and $\theta_{{\rm AI}} + 0.05$ respectively. 
    \item Students 6 and 7 (both AI-users) exhibit intermittency.
    The switch from stability to turbulent nonconvergence (or vice versa) is triggered, when an AI-user receives an ``abnormal'' score sequence, e.g.\ many consecutive scores satisfying $S_i(t) \approx 1$, or 
    their allies or opponents receive an abnormal score sequence and transmit its impact by peer pressure.
    \item Students 1, 3 (AI-users) and students 0, 2, 5 (AI-avoiders) exhibit turbulent nonconvergence.
    The fluctuations are more pronounced for AI-users, who observe $S_i(t)$ directly, as discussed in Section \ref{subsec: correct_allies_only}.
    In contrast, AI-avoiders fluctuate, when they are influenced by AI-users, who exhibit turbulent nonconvergence.
    \item Student 9 (AI-avoider) is an outlier; more than anyone else, they are governed by the specifics of their connections.
    Student 9 is an opponent of all the other students except student 3 (AI-user).
    Additionally, student 9 has a prior belief, that the AI-tool has low proficiency, with $\langle \theta \rangle |_{t = 0} \approx 0.38$, the lowest prior within the network. 
    Near the start of the simulation, for $ t \geq 47$, student 9's belief PDF satisfies $x_9(t, \theta > 0.3) = 0$ under the influence of antagonistic interactions. 
    Despite being allied with student 3, who has $x_3(t, \theta \leq 0.3) \neq 0$ and $x_3(t, \theta = 0.75 ) \neq 0$, we find $\Delta x_9 '(t + 1/2, \theta = 0.75) = 0$ as a result of the antagonistic interactions from opposing students 1, 2, 5, implying $x_9(t, \theta > 0.3) = 0$ for all $t < T$.
    We also observe that $x_9(t, \theta = 0.3)$ increases, as $t$ increases, and more scores $S_3(t)$ are observed by student 3, which updates student 9's opinion via Eq.\ \eqref{eq:xiprimed}.
    Hence, $\langle \theta \rangle$ increases as a result of an increase in $x_9(t, \theta = {0.3})$, as $t$ increases. 
\end{itemize}

For the sake of completeness, we run the experiment above on a Barab\'{a}si-Albert network as well, with $n = 10, m = 3$ and $k = 5$, performing 100 simulations with randomized priors.
Overall, we do not observe  any meaningful, qualitative difference in the students' long-term behavior.
We still  observe that AI-users and AI-avoiders achieve asymptotic learning.
Specifically, 249 AI-users achieve asymptotic learning, of which 156 (63\%) infer $\theta_{\rm AI}$ correctly, while 266 AI-avoiders achieve asymptotic learning, of which 141 (53\%) infer $\theta_{\rm AI}$ correctly.
Students who do not achieve asymptotic learning exhibit turbulent nonconvergence or intermittency. 
Behavior like that exhibited by Student 9 in Fig.\ \ref{fig:mixed} is not observed in the 100 simulations. 
However, it is hard to know whether this is because Student 9 is a rare outlier or because their behavior only occurs in a complete network. 
Answering this question and many others would require an efficient way to simulate a larger fraction of the combinatorially numerous mixed networks. 
Such an exercise lies outside the scope of this introductory paper and the computational resources at the authors' disposal.


\section{Disruption by partisans}
\label{sec:partisan_student}

We now extend the analysis in Section \ref{sec:allies} to treat partisans: obdurate agents, whose opinions do not change with time in response to independent observations or peer pressure but modify the opinions of other, persuadable agents through peer pressure.
Their obduracy may be sincere (e.g.\ rooted in psychology or ideology) or insincere (e.g.\ a tactic to disrupt a social network for political purposes by knowingly espousing a false opinion). 
Many published agent-based models treat partisans (who are sometimes termed zealots) as having deterministic and binary opinions, e.g.\ supporting or opposing a political party with zero uncertainty, as in the classic voter model \cite{mobilia_role_2007,mobilia_does_2003}.
By contrast, Ref.\ \cite{bu_discerning_2023} examines how partisans shape perceptions of media bias in a probabilistic, Bayesian framework, in which the partisans and agents have belief PDFs, which are multimodal in general, even though the partisans' PDFs are constant in time.
One interesting discovery is that a single partisan is enough to disrupt an arbitrary large network under certain plausible circumstances, either by persuading one or more agents to believe in a falsehood in the deterministic voter model \cite{mobilia_role_2007,mobilia_does_2003,mobilia_voting_2005} or by driving turbulent nonconvergence in the probabilistic Bayesian framework \cite{bu_discerning_2023}.

In this section, we introduce partisans into the model in Section \ref{sec:motivation}. 
As far as the authors are aware, this is the first time that partisans are investigated in the context of AI assistance in an educational setting.
In the latter context, obduracy is likely to be sincere, e.g.\ an unshakable ethical belief in the pedagogical value or otherwise of an AI tool, which translates (perhaps illogically) into an unshakable belief in the proficiency or otherwise of the tool\footnote{
    One may also imagine scenarios, where the obduracy is insincere, e.g.\ a disaffected student seeking to sow discord in the class.}.
It becomes important in an educational setting to distinguish between a partisan student and a partisan teacher. 
Both may be obdurate for similar psychological or ideological reasons, but the pattern of trust relationships in the network is different for students (one-to-one or one-to-few) and teachers (one-to-many), which affects the opinion dynamics\footnote{
    Arguably teacher-student trust relationships are more unidirectional than student-student trust relationships. To study this possibility, one may model the network as a directed graph, which lies outside the scope  of this paper.}.
We study the case of a partisan student and a partisan teacher in Sections \ref{subsec:student-partisan} and \ref{subsec:teacher-partisans} respectively. 
We copy the successful implementation strategy for partisans described in Section 2.2 of Ref.\ \cite{bu_discerning_2023}. 
In this paper, for the sake of simplicity and as a first pass, we only consider one partisan in the network and specialize to partisans who hold a single belief $\theta_{\rm p}$, i.e.\ $x_{\rm p}(t,\theta) = \delta(\theta-\theta_{\rm p})$. 

\subsection{Partisan student}
\label{subsec:student-partisan}
\begin{figure}[h!]
    \begin{subfigure}{0.5\textwidth}
        \includegraphics[width=\linewidth]{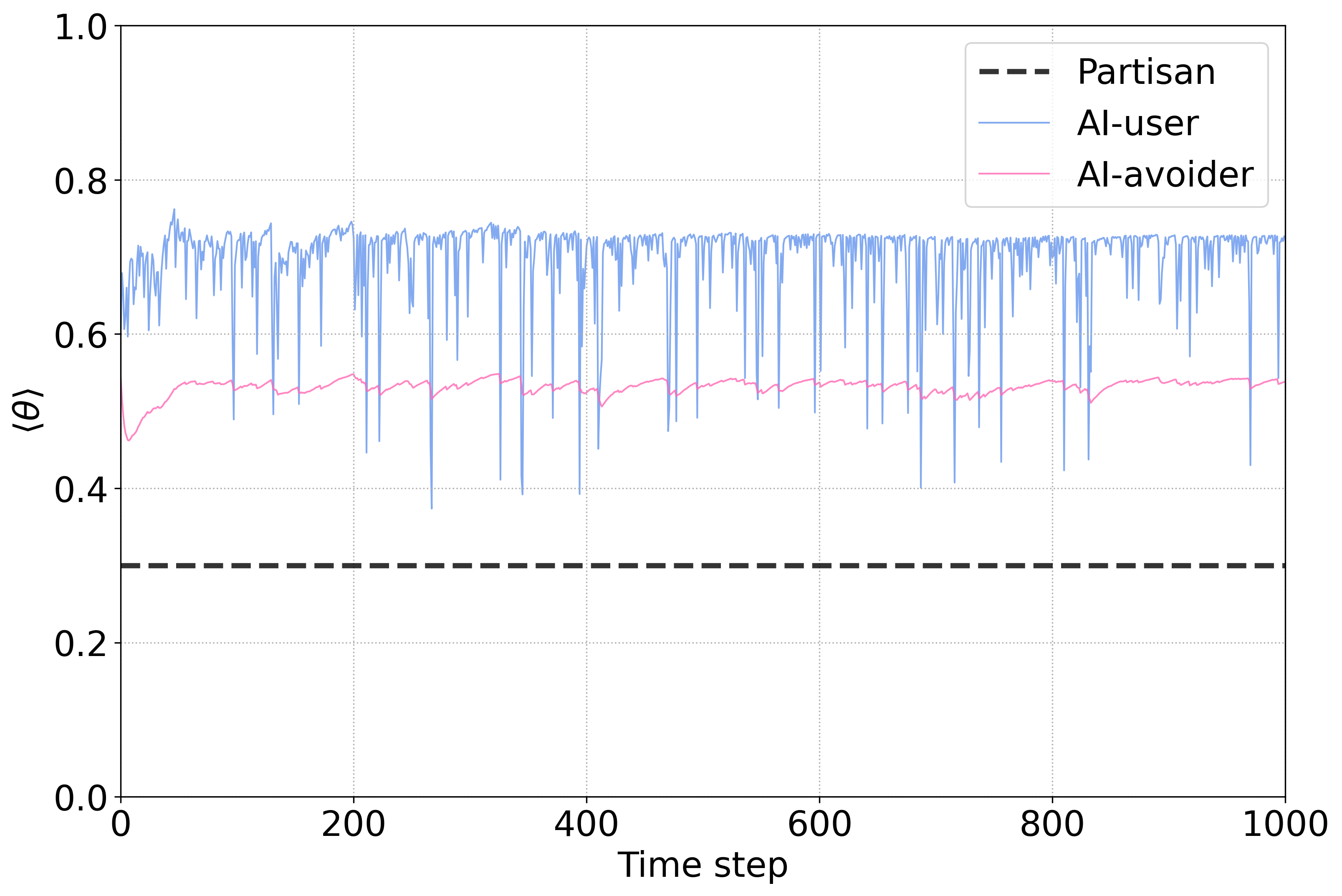}
    \end{subfigure}
    \hspace*{\fill}
    \begin{subfigure}{0.5\textwidth}
        \includegraphics[width=\linewidth]{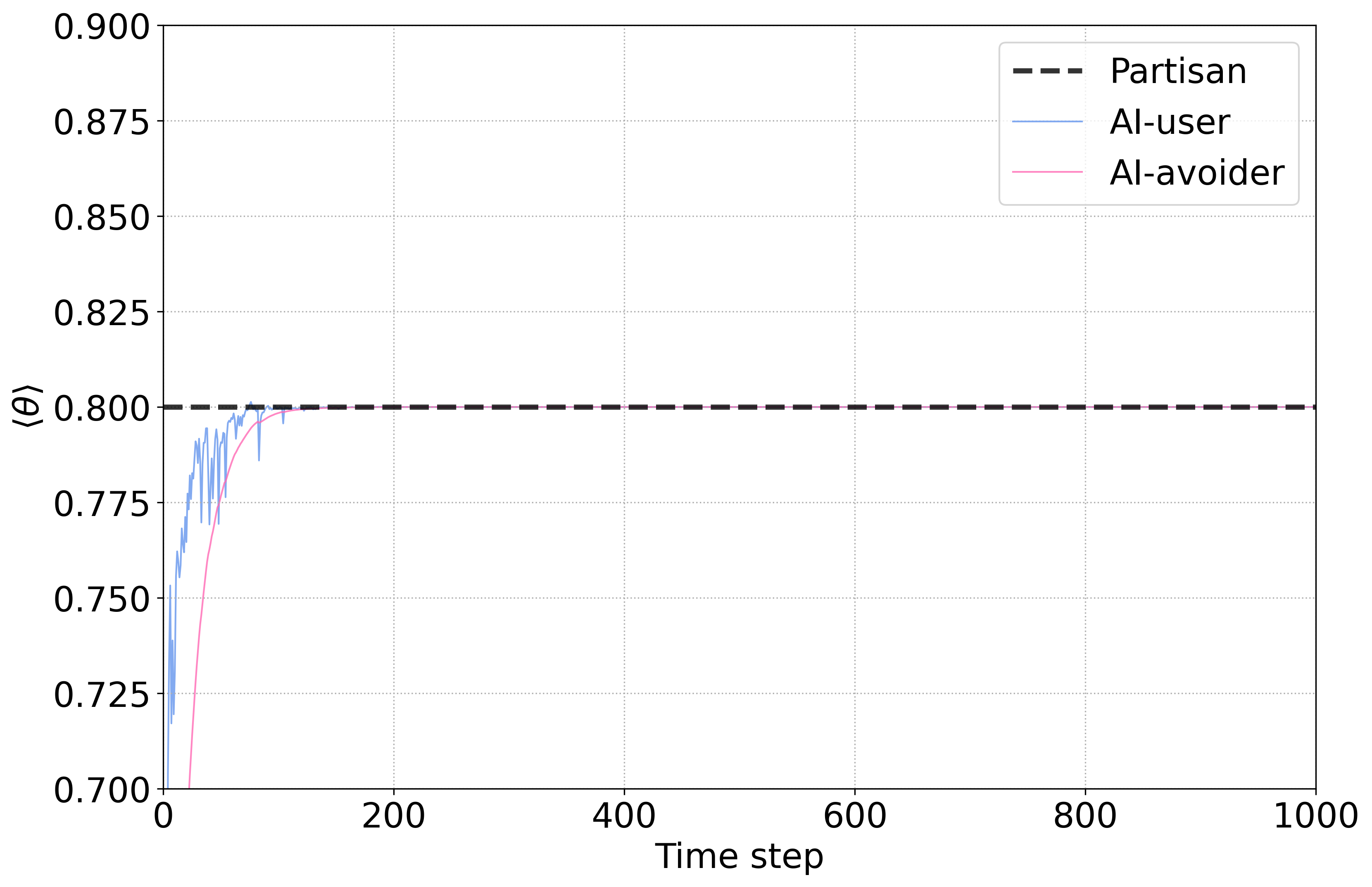}
    \end{subfigure}%
    \\
    \hspace*{\fill}  
    \begin{subfigure}{0.6\textwidth}
        \centering
        \includegraphics[width=\linewidth]{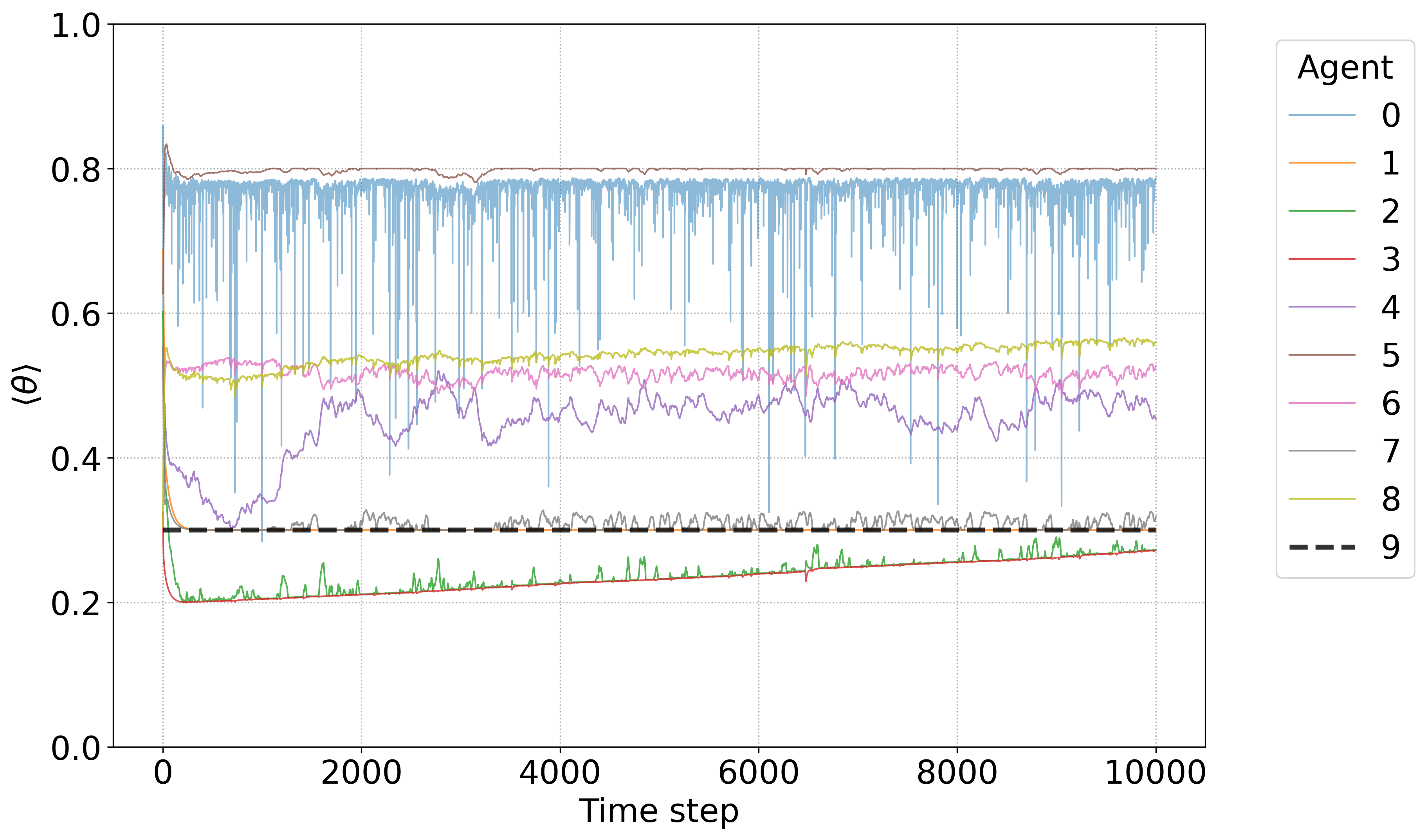}
    \end{subfigure} 
    \hspace*{\fill}  
    \caption{
        Disruption by a partisan student in an allies-only complete network (top panels) and a mixed complete network (bottom panel) with $n = 10$.
        (Top left.)
            Mean belief $\langle \theta \rangle$ versus time with one AI-user, one partisan, $\theta_{\rm p} =0.3$, and $\theta_{\rm AI}=0.8$.
            The AI-user (blue curve) and a representative AI-avoider (pink curve) exhibits turbulent nonconvergence.
        (Top right.)
            Same as top left, but with $\theta_{\rm p} = \theta_{\rm AI} = 0.8$, i.e.\ the partisan student is correct about the AI tool's proficiency. 
            The AI-user (blue curve) and a representative AI-avoider (pink curve) achieve asymptotic learning.
        (Bottom.)
            Mean belief $\langle \theta \rangle$ versus time in a representative simulation with one AI-user (student 0, blue curve) and one partisan with $\theta_{\rm p} = 0.3$ (student 9, black bold dashed curve).
            The eight AI-avoiders (students 1--8) are color-coded according to the legend.
            There are 25 pairs of allies and 20 pairs of opponents. 
            The long-term behavior depends sensitively on $A_{ij}$.
        }
    \label{fig:partisans}
\end{figure}

Let us first consider the scenario, where teachers are absent; all agents in the network are students.
Consider a complete network of $n=10$ allies, containing one partisan student and one AI-user.
We focus deliberately on a small class or study group ($n=10$) to compare directly with the results in Section \ref{sec:allies}; larger classes are discussed in Section \ref{sec:larger_networks}. 
In an introductory suite of tests, we assume that the partisan believes wrongly, that the AI tool is less proficient ($\theta_{\rm p} = 0.3$) than in reality ($\theta_{\rm AI} = 0.8$). 
The results are summarized in Fig.\ \ref{fig:partisans}.

The main result returned by the tests with $\theta_{\rm p} < \theta_{\rm AI}$ is that all AI-users and AI-avoiders exhibit turbulent nonconvergence until $t = T$, even though there is only one partisan, and all the students are allies, as shown in the top left panel in Fig.\ \ref{fig:partisans}.
Each student has a bimodal belief PDF satisfying $x_i(t, \theta = \theta_{\rm AI}) \neq 0$ and $x_i(t, \theta = \theta_{\rm p}) \neq 0$.
The mean belief $\langle \theta \rangle$ fluctuates for all $t$, as the probability weights of the two modes fluctuates; for example, the time-averaged standard deviations of $\langle \theta \rangle(t)$ for the plotted AI-user and AI-avoider are $5.8 \times 10^{-2}$ and $8.3 \times 10^{-3}$ respectively for $0.5T \leq t \leq T$. 
This phenomenon agrees with the counterintuitive finding in Ref.\ \cite{bu_discerning_2023}, that a single partisan destabilizes an allies-only network regardless of its size, stopping everyone from achieving asymptotic learning. 
In the educational context, this result raises a challenging ethical quandary. 
It is healthy to foster a climate of mutual trust within a class for many reasons unrelated to AI assistance, and it is healthy to encourage students to form their own views and express them confidently.
Yet a class with universally positive trust relationships is disrupted by even one obdurate, misinformed student, even if their obduracy is sincere and independent evidence about the truth [here, $S_i(t)$] is available freely and on repeated occasions to the whole class. 
AI-users and AI-avoiders exhibit the same long-term behavior. 
They are both pulled towards $\theta_{\rm p}$ by the partisan though Eq.\ \eqref{eq:xiprimed}.
They are also influenced by $S_i(t)$, which pulls students towards $\theta_{\rm AI}$.
AI-users observe $S_i(t)$ directly, while AI-avoiders feel the pull of $S_i(t)$ indirectly, via the belief PDFs of allied AI-users.
Additionally, we observe that AI-avoiders reach consensus among themselves, and always have $x_i(t, \theta = \theta_{\rm p}) > x_i(t, \theta = \theta_{\rm AI})$ after consensus is reached.
AI-users on the other hand do not reach consensus, as a result of the perturbations from $S_i(t)$.

In a second suite of tests, we assume that the partisan is correct about the proficiency of the AI tool.
For $\theta_{\rm p} = \theta_{\rm AI}$, both AI-users and AI-avoiders reach asymptotic learning faster than they do in a network without partisans. 
We run 100 simulations in a complete allies-only network with $n = 10, k = 1$, and one partisan.
The mean belief $\langle \theta \rangle$ of the partisan, the AI-user, and one AI-avoider for a representative simulation evolve as displayed in the top right panel of Fig.\ \ref{fig:partisans}.
The selected AI-avoider is representative of every AI-avoider, as they reach consensus among themselves for the same reason discussed in Section \ref{subsec: correct_allies_only}.
We no longer observe the fluctuations that are observed in networks without partisans (see Section \ref{subsec: correct_allies_only})\footnote{
    The behavior does not depend on the exact value of $\theta_{\rm AI}$ and $\theta_{\rm p}$ but is different for $\theta_{\rm AI} = \theta_{\rm p}$ and $\theta_{\rm AI} \neq \theta_{\rm p}$.
}. 
Under the influence of a partisan, the AI-user evolves to $x_i(t, \theta = \theta_{\rm AI}) = 1$; as Eq.\ \eqref{eq:updatefirsthalf_ai} is multiplicative, the Gaussian fluctuations in $S_i(t)$ ``zero out''. 
Once the AI-user's belief stops fluctuating, so do the AI-avoiders' beliefs. 
The trend where AI-users achieve asymptotic learning later than AI-avoiders is still observed, when the partisan is in the network.
For example, for $k = 1$, the $t_{\rm a}$ histogram peaks at $t_{\rm a} \approx 43$ and 110 respectively for AI-avoiders and AI-users. 
However, we do not observe that the modal $t_{\rm a}$ decreases monotonically with $k$, because the influence of the correct partisan dominates the Gaussian fluctuations and accelerates asymptotic learning, consistent with the top right panel of Fig.\ \ref{fig:partisans}.

The results in the previous paragraphs in Section \ref{subsec:student-partisan} hold qualitatively, even when the network is not complete. 
We demonstrate such qualitative agreement for Barab\'{a}si-Albert networks with $n=10$ and $m=3$ in \ref{sec:incomplete}.

We now consider a mixed complete network with $n = 10, k = 1$, and one partisan with $0.3 = \theta_{\rm p} \neq \theta_{\rm AI} = 0.8$\footnote{
    Opponents-only networks containing one partisan student behave exactly the same in the long term as networks without a partisan, as in Section \ref{subsec: incorrect_opponent_only}, because a partisan effectively plays the same role as any other opponent, who quickly reaches asymptotic learning and settles on a correct or incorrect belief. 
}.
Here the picture grows complicated. 
Enumerating every possible scenario falls outside the scope of this introductory paper. 
As a rule, though, we find that the long-term behavior of students in a mixed network depends sensitively on the relationship pattern, especially the relationship between the AI-user and partisan.
When the AI-user and the partisan are direct allies, or indirect allies joined by a student who is allied to both, the AI-user exhibits turbulent nonconvergence, vacillating between $\theta_{\rm AI}$ and $\theta_{\rm p}$.
The bottom panel in Fig.\ \ref{fig:partisans} graphs $\langle \theta \rangle$ versus time in a representative simulation, where the AI-user is connected directly to the partisan.
Student 0 is the AI-user (blue curve) and student 9 is the partisan (bold, black, dashed curve). 
Student 5 (brown curve) is allied with the AI-user, correctly infers $\theta_{\rm AI}$, and is repelled from $\theta_{\rm p}$ by the opposing partisan.
By contrast, students 1 and 7 (orange and grey curves respectively), who oppose the AI-user but are allies of the partisan, infer $\theta_{\rm p}$, with $t_{\rm a, 1} = 29$ and $t_{\rm a, 7} = 41$.
Students 2 and 3 (green and red curves respectively) behave like student 9 in the mixed network discussed in Section \ref{subsec: turbulent_nonconvergence}.
As student 2 is allied with the AI-user, their belief vacillates under the influence of $S_0(t)$ indirectly, through the AI-user's PDF.
Students 4, 6 and 8 (purple, pink, lime curves respectively) exhibit turbulent nonconvergence, under the influence of the allied AI-user and partisan. 
The antagonistic interaction between students 4 and 6 can be clearly seen: after the initial transient, $\langle\theta\rangle$ increases for student 4 yet decreases for student 6.

As noted above, many other scenarios are possible in a mixed network, whose enumeration falls outside the scope of this paper. 
As one example, when the AI-user and partisan are opponents, we sometimes observe the same long-term behavior as seen in the top panel of Fig.\ \ref{fig:diff_t_A}. 
Alternatively, we observe instances where all students achieve asymptotic learning. 
It will be interesting to categorize the long-term behavior by the network topology, or by the theory of social balance \cite{heider_attitudes_1946,cartwright_structural_1956}, in future work. 

\subsection{Partisan teacher}
\label{subsec:teacher-partisans}


We now introduce a partisan teacher into the system.
We assume that the teacher is unaffected by the students' beliefs but is allied with, and therefore influences positively, every student in the class. 
The scenario is plausible as a first approximation, given the authority enjoyed by a teacher in certain educational settings, but other plausible scenarios exist of course. 
The teacher may be an AI-skeptic, who fears that AI tools encourage lazy thinking and therefore preaches that AI tools display low proficiency ($\theta_{\rm p} < \theta_{\rm AI})$, to discourage students from using them. 
Alternatively, the teacher may be an AI-promoter, who exaggerates the proficiency of AI tools ($\theta_{\rm p} > \theta_{\rm AI}$) either sincerely or manipulatively to encourage students to use them.
In this section, as a concrete example, we construct a complete network with $0 \leq i < n=10$ students with mixed allegiances, then add a teacher node ($i =10$) allied with every student. 
The associated adjacency matrix is given by
\begin{align}\label{eq:adj_mat}
    A_{ij} = \begin{pmatrix}
        0  & -1  & +1  & +1  & -1  & -1  & +1  & -1  & -1  & +1  & +1 \\
        -1  &  0  & -1  & +1  & +1  & +1  & -1  & +1  & -1  & -1  & +1 \\
        +1  & -1  &  0  & -1  & -1  & +1  & -1  & -1  & +1  & -1  & +1 \\
        +1  & +1  & -1  &  0  & +1  & -1  & +1  & -1  & +1  & +1  & +1 \\
        -1  & +1  & -1  & +1  &  0  & +1  & +1  & +1  & +1  & +1  & +1 \\
        -1  & +1  & +1  & -1  & +1  &  0  & -1  & +1  & -1  & +1  & +1 \\
        +1  & -1  & -1  & +1  & +1  & -1  &  0  & +1  & +1  & +1  & +1 \\
        -1  & +1  & -1  & -1  & +1  & +1  & +1  &  0  & +1  & -1  & +1 \\
        -1  & -1  & +1  & +1  & +1  & -1  & +1  & +1  &  0  & -1  & +1 \\
        +1  & -1  & -1  & +1  & +1  & +1  & +1  & -1  & -1  &  0  & +1 \\
        +1  & +1  & +1  & +1  & +1  & +1  & +1  & +1  & +1  & +1  &  0 \\
        \end{pmatrix}.
\end{align}
Student 0 is an AI-user; the other nine students are AI-avoiders.

\begin{figure}[h!]
    \begin{subfigure}{0.5\textwidth}
        \includegraphics[width=\linewidth]{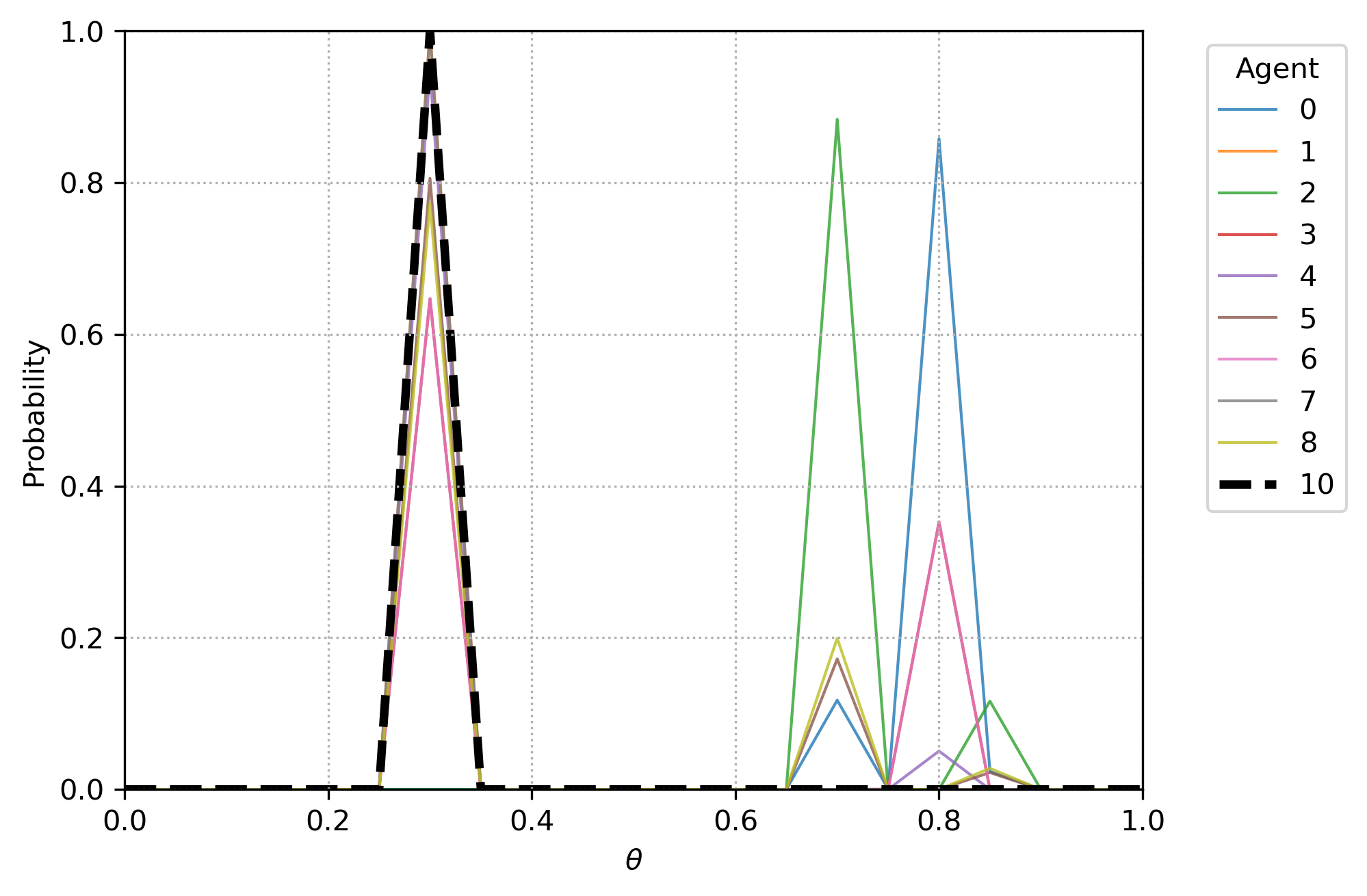}
    \end{subfigure}
    \hspace*{\fill}
    \begin{subfigure}{0.5\textwidth}
        \includegraphics[width=\linewidth]{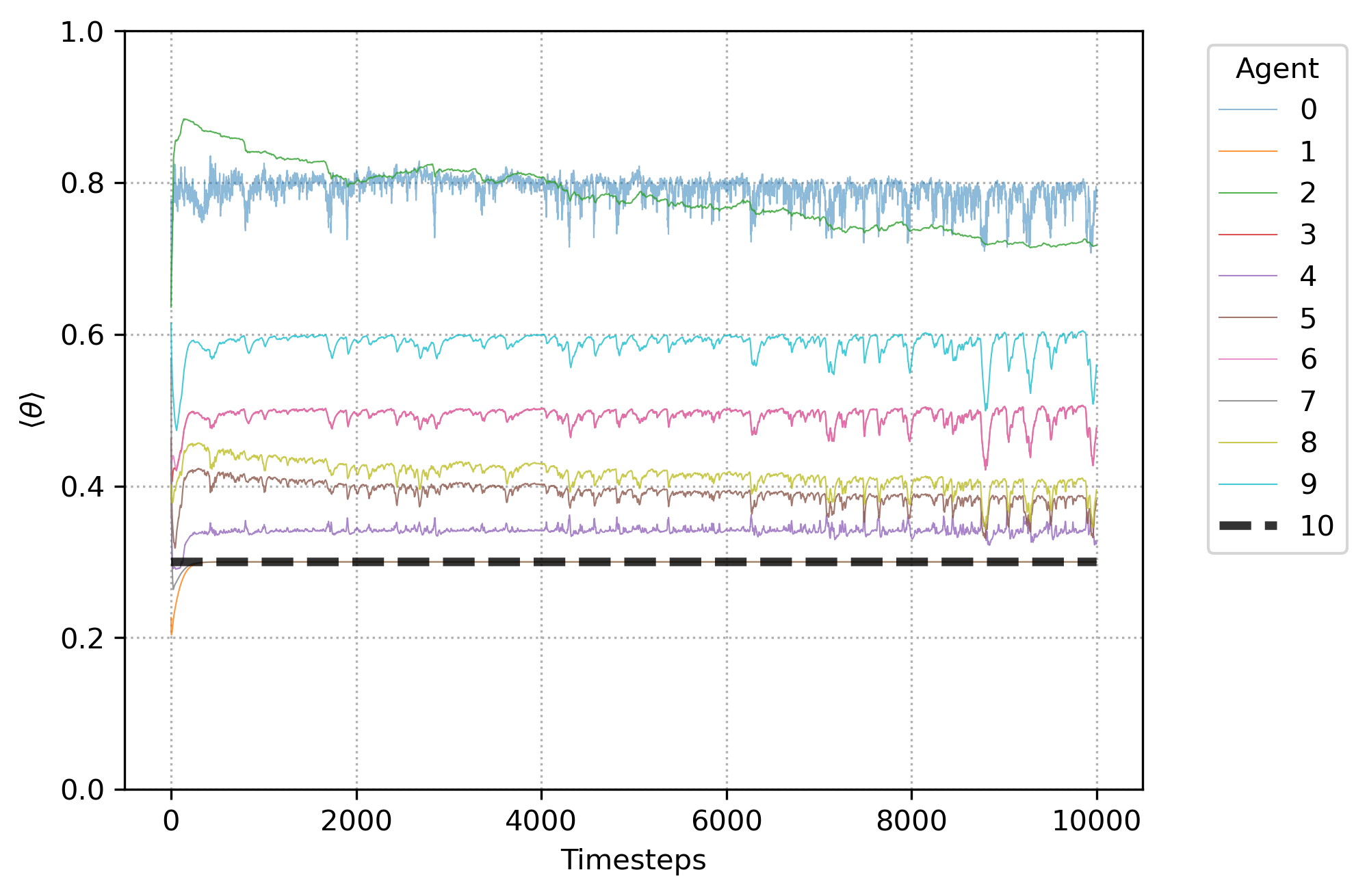}
    \end{subfigure}%
    \\
    \begin{subfigure}{0.5\textwidth}
        \includegraphics[width=\linewidth]{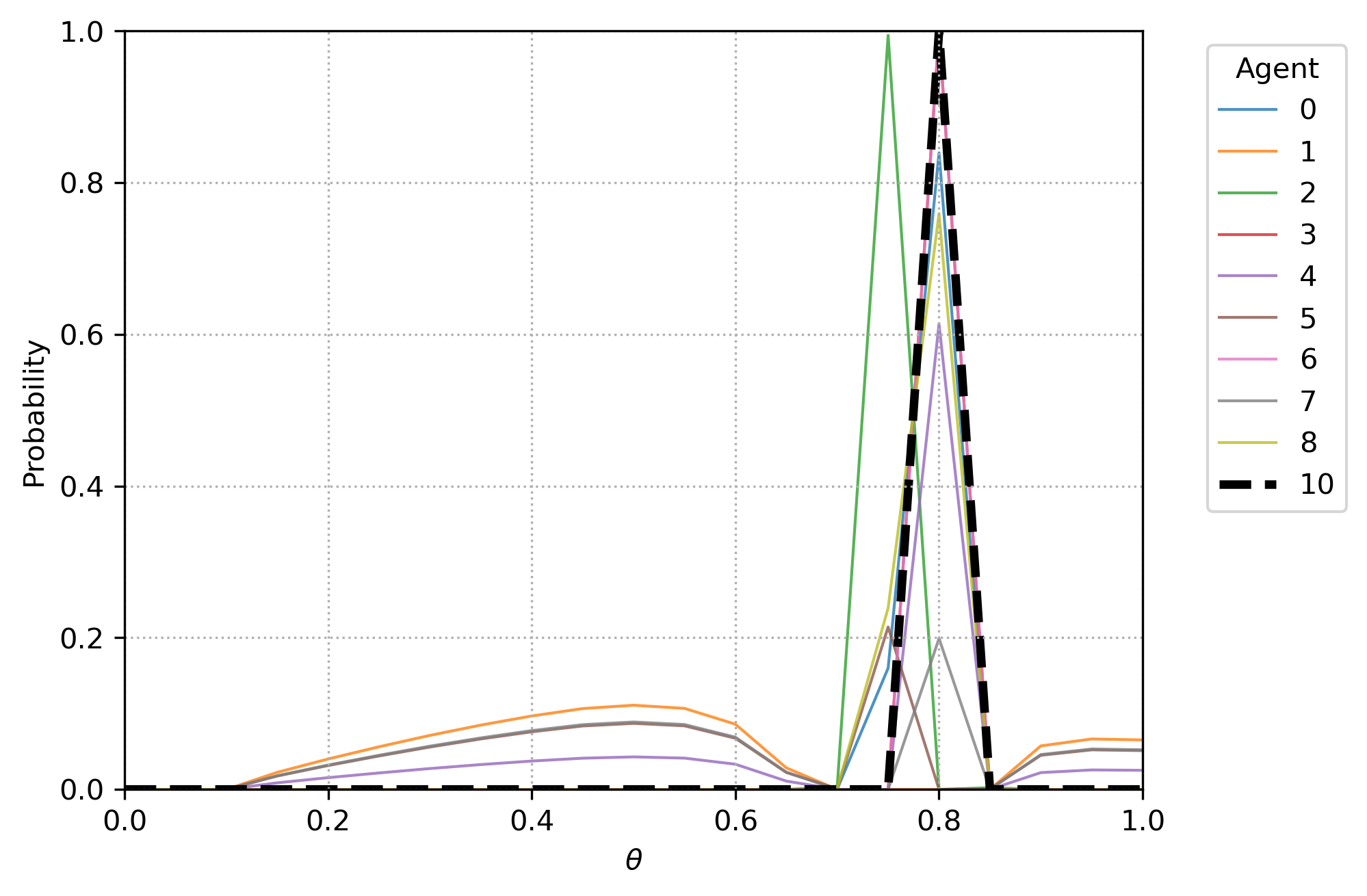}
    \end{subfigure}%
    \hspace*{\fill}  
    \begin{subfigure}{0.5\textwidth}
        \centering
        \includegraphics[width=\linewidth]{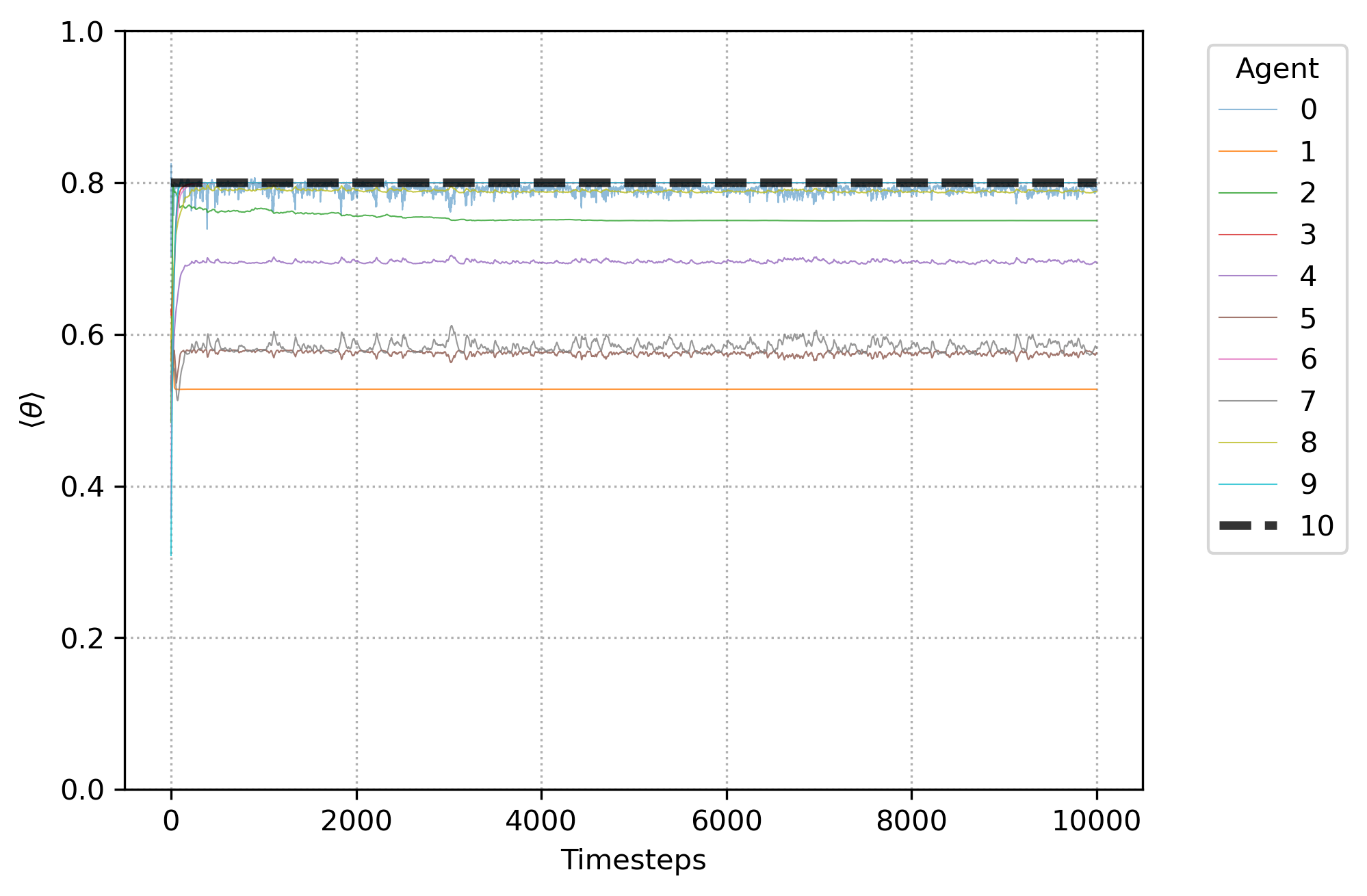}
    \end{subfigure} 
    \hspace*{\fill}  
    \caption{
        Disruption by a partisan teacher in a mixed complete teacher-student network with $n = 11$.
        The partisan teacher (bold, black, dashed curve) is allied with every student.
        Student 0 is the only AI-user (blue solid curve).
        The weighted adjacency matrix is defined by Eq.\ \eqref{eq:adj_mat}.
        The top and bottom rows correspond to $\theta_{\rm p} = 0.3 \neq 0.8 = \theta_{\rm AI}$ and $\theta_{\rm p} = 0.8 = \theta_{\rm AI}$ respectively.
        The left column displays a snapshot of the belief PDF at $t = 10^3$ for all students and the partisan teacher. 
        The right column displays the evolution of the mean $\langle\theta\rangle$ of the belief PDF for each agent.
        Students 1 and 7 achieve asymptotic learning, reaching consensus with the partisan teacher when the teacher is an AI-skeptic (top panels), while all other students exhibits turbulent nonconvergence.
        When the teacher is correct about the proficiency of the AI tool, 2925 out of $10^4$ students spanning $10^3$ simulations infer the AI tool's proficiency correctly and reach asymptotic learning.
            }
    \label{fig:partisan_teacher}
\end{figure}

We start by setting $\theta_{\rm p} = 0.3$ and $\theta_{\rm AI} = 0.8$.
We run $10^3$ simulations \footnote{
    We run $10^3$ simulations in this experiment, unlike in other sections where only $10^2$ simulations are conducted. 
    We can afford to do so because we only investigate one network configuration.
    Other sections investigate multiple network configurations (different $k$ values, for example), which raise storage constraints.} 
on the network defined by Eq.\ \eqref{eq:adj_mat} with randomized initial priors and $S_i(t)$. 
The top two panels of Fig.\ \ref{fig:partisan_teacher} show a snapshot, for a representative simulation, of the belief PDF of every agent at the end of the simulation (top left panel) as well as the evolution of $\langle \theta \rangle$ for every agent (top right panel).
The bold, black, dashed curve represents the  partisan teacher, who is an AI-skeptic in Fig.\ \ref{fig:partisan_teacher}, and the colored curves represent students. 
The AI-user exhibits turbulent nonconvergence with a bimodal belief PDF, as shown by the blue curve in the top left panel in Fig.\ \ref{fig:partisan_teacher}, with $x_{\rm AI-user}(t, \theta = 0.7) \neq 0$ and $x_{\rm AI-user}(t, \theta = \theta_{\rm AI}) \neq 0$.
This behavior occurs in every simulation; it is not specific to the representative simulation in Fig.\ \ref{fig:partisan_teacher}.
The AI-user's opinion is influenced only by the AI scores and the opinions of allied agents, not the partisan teacher. 
The teacher is outweighed by many students who oppose the AI-user. 
Under the positive influence of the partisan teacher, the AI-user's opponents have $x_i(t,\theta_{\rm p}) \neq 0$.
That is, one has
$$\left| \sum_{A_{0j} = -1} x_j'(t + 1/2, \theta = \theta_{\rm p}) - x_0'(t + 1/2, \theta = \theta_{\rm p}) \right| \geq \left| \sum_{A_{0j} = 1} x_j'(t + 1/2, \theta = \theta_{\rm p}) - x_0'(t + 1/2, \theta = \theta_{\rm p}) \right|,$$
implying $\Delta x_0'(t + 1/2, \theta_{\rm p }) \leq 0$. 
Similarly, student 2 (green curve) has $x_2(t, \theta_{\rm p}) = 0$ despite being allied with the partisan teacher.
The teacher's preference for $\theta=\theta_{\rm p}$ is outweighed by the antagonistic interaction between student 2 and six opponents, who all have $x_i(t, \theta_{\rm p}) \neq 0$.

Students who are opponents of the AI-user do not gain information about the AI tool's proficiency directly; they are influenced by common non-partisan (CNP) allies of the AI-user\footnote{
    Partisans block the information flow in the network.
    If students $i$ and $j$ are allies with the same partisan but are not connected with each other, $j$ cannot obtain information about $x_i(t,\theta)$ through the partisan, and vice-versa.}. 
We find that, for students who oppose the AI-user, students with fewer CNP allies are influenced less by the AI-user than students with more CNP allies.
We observe that students 1 (orange curve) and 7 (grey curve) who oppose the AI-user and have only one AI-user CNP ally (students 3 and 6 respectively) always reach asymptotic learning at $\theta_{\rm p}$, regardless of their initial prior beliefs and $S_i(t)$.
Students 4 (purple curve), 5 (brown curve) and 8 (lime curve) also oppose the AI-user, but by contrast they have more than one AI-user CNP ally.
They exhibit turbulent nonconvergence under the influence of the teacher and the AI-user.

We now set $\theta_{\rm p} = \theta_{\rm AI}$, so that the partisan teacher is correct about the AI tool's proficiency. 
As above, we run $10^3$ simulations on the network defined by Eq.\ \eqref{eq:adj_mat} with randomized initial priors and $S_i(t)$.
The bottom two panels in Fig.\ \ref{fig:partisan_teacher} show a snapshot of the belief PDF at the end of the simulation (bottom left panel) and the evolution of $\langle\theta\rangle$ for every agent in a representative simulation (bottom right panel).
Across all $10^3$ simulations, we observe that 2925 students (out of $10^4$ students) achieve asymptotic learning and correctly infer the AI tool's proficiency.
Students 3 (red curve), 6 (pink curve) and 9 (cyan curve) reach asymptotic learning at $\theta_{\rm AI}$ more frequently, doing so in 942, 945, 945 out of the $10^3$ simulations respectively.
Students 0, 1, 2, 7, 8 reach asymptotic learning at $\theta_{\rm AI}$ rarely, i.e.\ in $\approx 20$ out of $10^3$ simulations. 
Student 4 and 5 are not observed to achieve asymptotic learning at all in this batch of $10^3$ simulations. 
The behavior of students when the partisan teacher is correct about the AI tool's proficiency depends sensitively on the student's prior beliefs and the AI score, unlike the previous experiment where the partisan teacher has $\theta_{\rm p} = 0.3 \neq \theta_{\rm AI}$.
Surprisingly, students are more likely to be uncertain, when the partisan teacher is correct about the AI tool's proficiency; for example, $x_1(t,\theta)$, $x_4(t,\theta)$, and $x_7(t,\theta)$ are non-zero at multiple $\theta$ values in the bottom left panel of Fig.\ \ref{fig:partisan_teacher}. 
In contrast, for $\theta_{\rm p} = 0.3$, we observe that $x_1(t,\theta)$, $x_4(t,\theta)$, and $x_7(t,\theta)$ are either unimodal or bimodal, as in the top left panel of Fig.\ \ref{fig:partisan_teacher}.
This is because the AI-user cancels the influence of the partisan teacher, when the student  is an ally of the teacher and an opponent of the AI-user.

\section{Larger networks}
\label{sec:larger_networks}
The investigations in Sections \ref{sec:allies} and \ref{sec:partisan_student} can be extended in several ways. 
As one example, Sections \ref{sec:allies} and \ref{sec:partisan_student} focus deliberately on relatively small classes ($n \sim 10$), such as those found in some high schools or postgraduate university degrees, where trust relationships (whether positive or negative) are dense. 
It is interesting to ask whether the behavior observed in small classes is replicated in larger, sparsely connected classes ($n \gtrsim 10^2$), examples of which include cross-disciplinary service subjects in the first undergraduate year of university (e.g.\ introductory mathematics taken by science, engineering, economics, and design majors).
A thorough analysis of larger classes is expensive computationally and is postponed to a forthcoming paper.
Nevertheless, to give a flavor of what is possible, we conduct a reduced set of preliminary experiments in this section, to check whether the key findings in Sections \ref{sec:allies} and \ref{sec:partisan_student} generalize to larger networks.
Specifically, we check the following questions:
\begin{itemize}
\item In larger allies-only networks, do AI-avoiders still achieve asymptotic learning faster than AI-users, as observed in Fig.\ \ref{fig:diff_t_A}?
\item In larger opponents-only networks, does a minority of students still infer $\theta_{\rm AI}$, with AI-users retaining an advantage over AI-avoiders, as observed in Section \ref{subsec: incorrect_opponent_only}?
\item In larger allies-only networks, does a single partisan still induce turbulent nonconvergence across the network, as observed in Fig.\ \ref{fig:partisans}?
\item In larger student-teacher networks, is the influence of the teacher more or less pronounced than in smaller networks?
\end{itemize}
In what follows, we simulate networks with $10^2 \leq n \leq 10^3$, subject to computational constraints.\footnote{
    The simulations in this paper are not time-bound; a run with $n = 10^3$ takes $\sim$10 minutes on a standard personal computer.
    However, storage is a bottleneck: a simulation with $n = 200$ and $T = 10^4$ requires $\sim$300 MB for analysis, and the storage required scales as $O(n^2)$.
    An ensemble of  many simulations with $n = 10^3$ consumes terabytes.
}
The simulated networks are allies- or -opponents-only (except for a brief mention of mixed networks in Section \ref{subsec:partisan_teacher_larger}). 
Large mixed networks involve too many permutations to be studied systematically with the computational resources at  the authors' disposal.

\subsection{Correctly inferring $\theta_{\rm AI}$ in allies-only networks}
\label{subsec:allies_larger}
\begin{figure}[h!]
    \centering
    \begin{subfigure}{0.45\textwidth}
        \includegraphics[width=\linewidth]{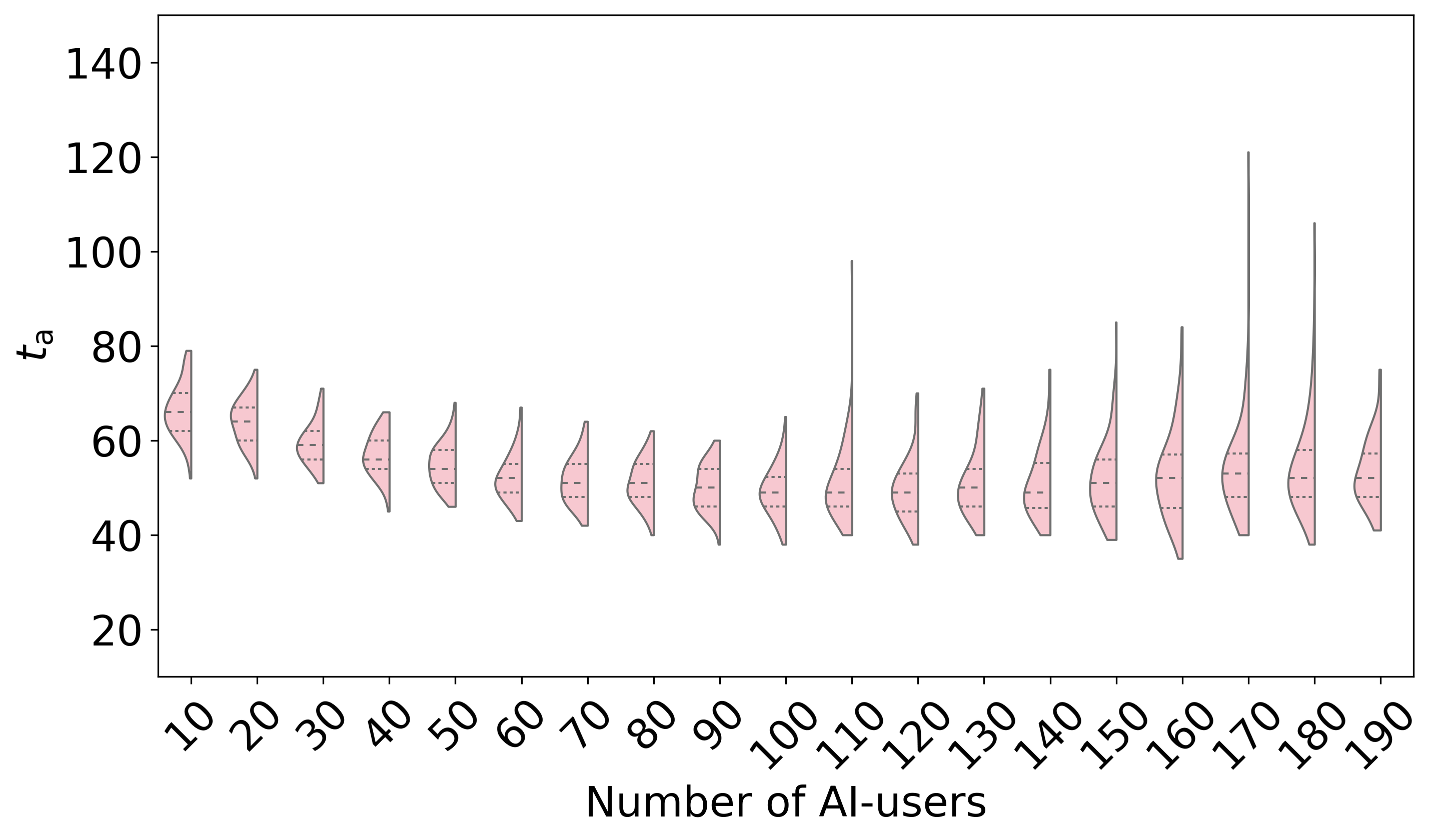}
    \end{subfigure}  
    \begin{subfigure}{0.45\textwidth}
        \includegraphics[width=\linewidth]{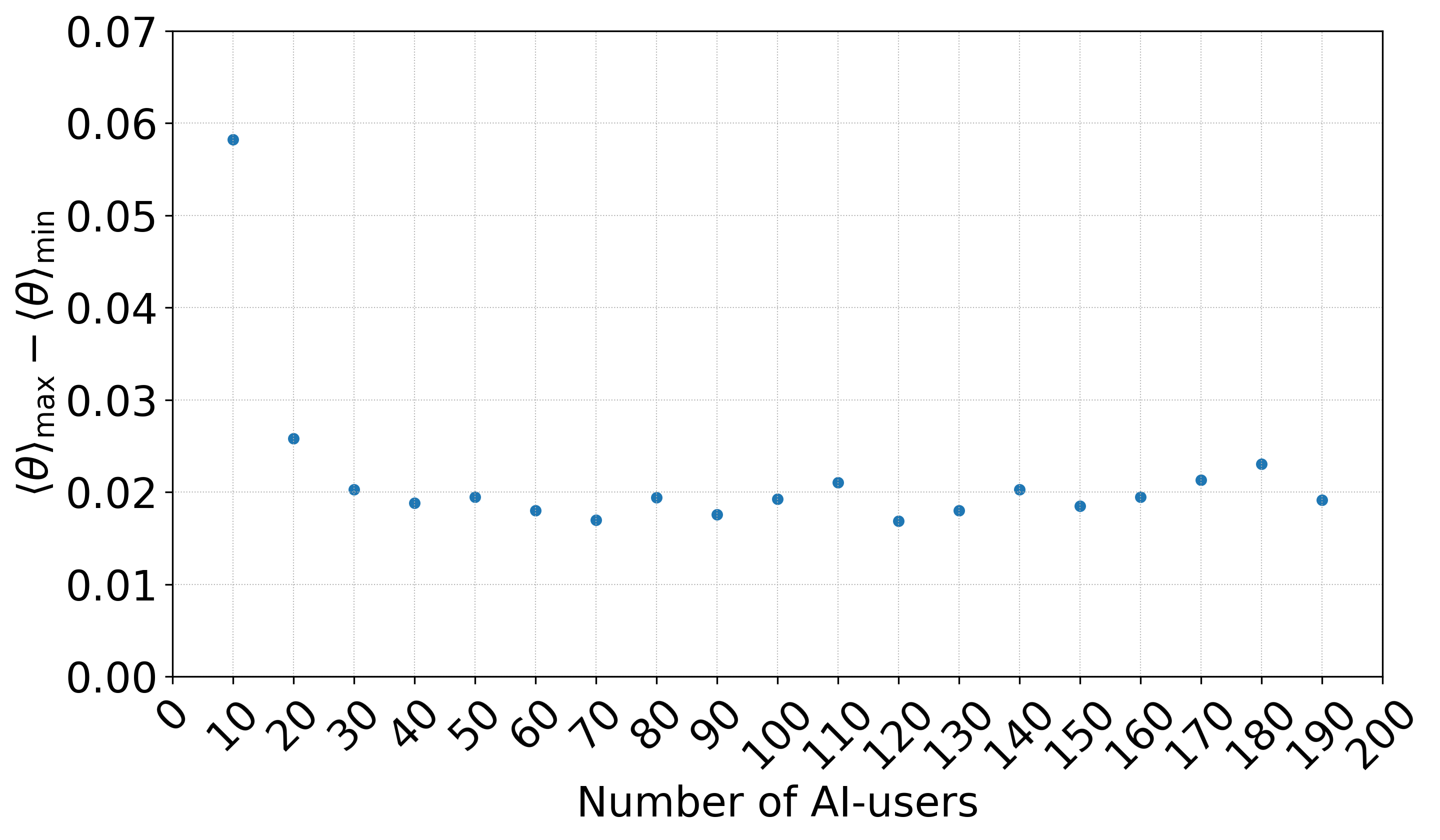}
    \end{subfigure} 
    \caption{
        Correctly inferring $\theta_{\rm AI}$ in a larger allies-only network with $n=200$, to be compared with the smaller network with $n=10$ in Fig. \ref{fig:diff_t_A}.
        (Left.) 
        Distribution of asymptotic learning time, $t_{\rm a}$, as a function of the number of AI-users, $k$, copying the violin-plot format in the lower panels of Fig.\ \ref{fig:diff_t_A}.
        We do not plot $t_{\rm a}$ for AI-avoiders, as all AI-avoiders in all simulations fail to achieve asymptotic learning by the end of the run (i.e.\ at $t=T=10^4$).
        (Right.)
        Spread of mean beliefs $\langle \theta \rangle_{\rm max} -\langle \theta \rangle_{\rm min}$ for an arbitrary, representative AI-avoider as a function of $k$, for a single, representative simulation per $k$ value.
        $\langle\theta\rangle_{\rm max}$ ($\langle\theta\rangle_{\rm min}$) equals the maximum (minimum) of $\langle\theta\rangle$ for an AI-avoider evaluated temporally, i.e.\ for $10^3 \leq t \leq 10^4$ to minimize the memory of the prior at $t=0$.
        }
    \label{fig:larger_allies_only}
\end{figure}

In larger allies-only complete networks, the counterintuitive behavior persists, wherein AI-avoiders correctly infer $\theta_{\rm AI}$ and achieve asymptotic learning faster than AI-users. 
By way of example, we consider complete allies-only networks with $n=200$ and $ k = 10, 20, \ldots, 190$, running 100 simulations for each $k$ value. 
The results are plotted in Fig.\ \ref{fig:larger_allies_only} in a format that resembles the bottom panels of Fig.\ \ref{fig:diff_t_A}.
AI-avoiders reach consensus among themselves for $t\gtrsim 20$.
The asymptotic learning time $t_{\rm a}$ for AI-avoiders depends weakly and non-monotonically on $k$, for the reasons explained in Section \ref{subsec: correct_allies_only}. 
Specifically, in the left panel of Fig.\ \ref{fig:larger_allies_only}, we find $49.3 \leq \overline{t_{\rm a}} \leq 66.7$ for $10 \leq k \leq 190$, where an overbar denotes an average over all AI-avoiders in the network.
AI-users fail to reach asymptotic learning before the end of the run ($t\leq T = 10^4$), as they experience directly the Gaussian fluctuations in $S_i(t)$. 
The right panel of Fig.\ \ref{fig:larger_allies_only} displays the spread in the beliefs of AI-avoiders. 
For the complete allies-only network with $n=200$, we run one representative for each $k$ value in steps of 10 and plot $\langle \theta \rangle_{\rm max} - \langle \theta \rangle_{\rm min}$ for a single, arbitrary AI-avoider, with $\langle \theta \rangle = \int_0^1 d \theta' \theta' x_i(t, \theta')$ for the $i$-th student as defined in Section \ref{subsec: correct_allies_only}. 
The subscripts max and min denote the maximum and minimum $\langle \theta \rangle$ achieved over time, i.e.\ for $10^3 \leq t \leq 10^4$ to minimize the memory of the prior at $t=0$.
As all AI-avoiders reach consensus, one AI-avoider is representative of their group behavior. 
For $k \lesssim 40$, the spread $\langle \theta \rangle_{\rm max} - \langle \theta \rangle_{\rm min}$ decreases with increasing $k$.
As AI-users gain more information about the AI tool through their neighbors’ beliefs, they are more certain about the AI-tool's proficiency.
For $k \gtrsim 40$, the spread depends weakly on $k$. 
The spread persists, because for large $k$ the average of the neighbors’ beliefs concentrates around $\theta_{\rm AI}$, while $S_i(t)$ continues to fluctuate, sustaining uncertainty in $x_i(t,\theta)$.

\subsection{Incorrectly inferring the AI's proficiency in opponents-only networks}
\label{subsec:larger_opponents}

\begin{figure}[h!]
    \begin{subfigure}{0.5\textwidth}        
        \includegraphics[width=\linewidth]{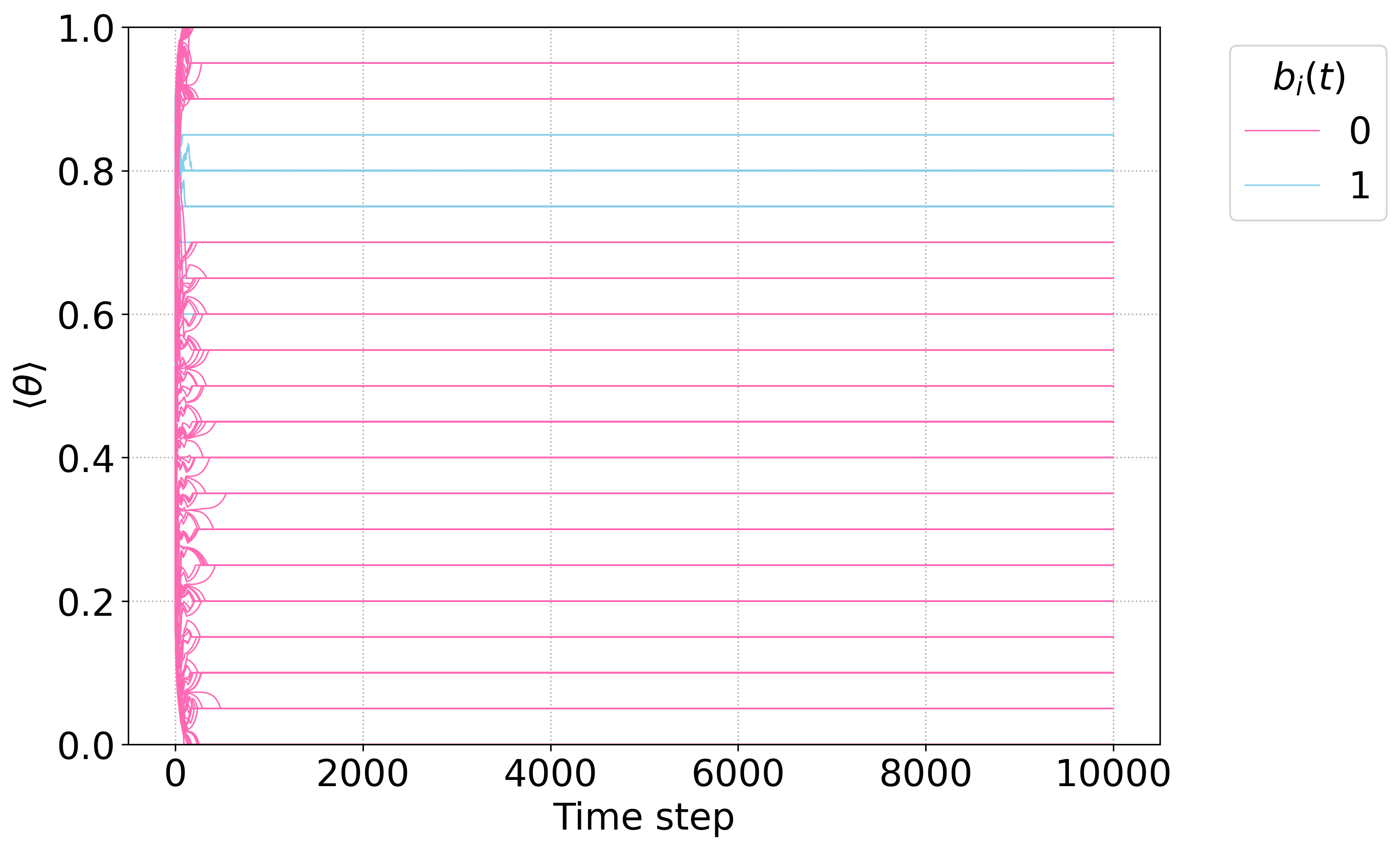}
    \end{subfigure}
    \begin{subfigure}{0.5\textwidth}        
        \includegraphics[width=\linewidth]{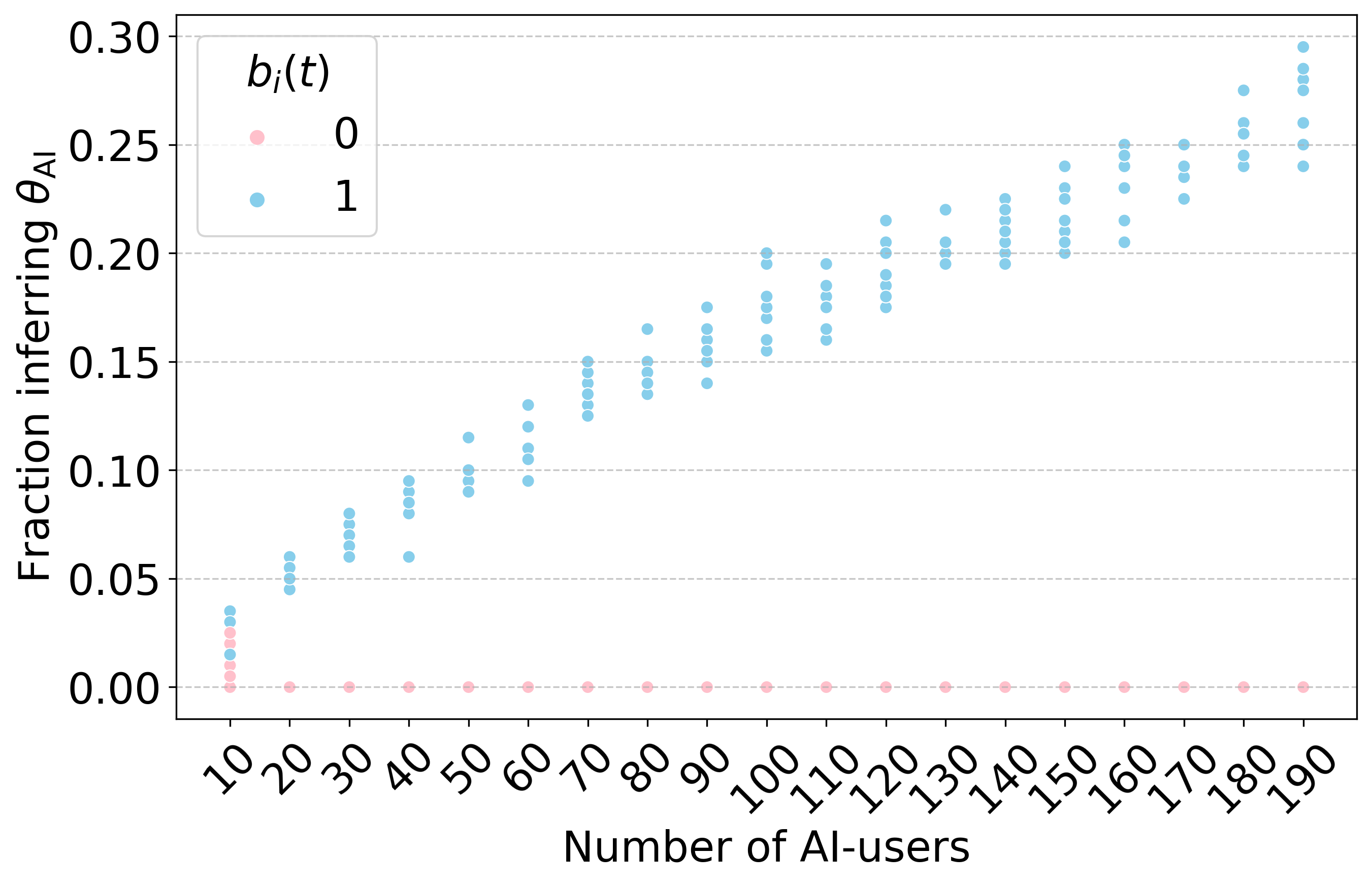}
    \end{subfigure}
    \caption{
        Incorrectly inferring $\theta_{\rm AI}$ in a large opponents-only complete network with $n=200$.
        (Left.)
        Same as the top left panel in Fig.\ \ref{fig:opponent} for $k = 40$. 
        Mean belief $\langle \theta \rangle$ versus time $t$ for AI-users (blue curves) and AI-avoiders (pink curves). 
        19 out of 40 AI-users and zero AI-avoiders infer $\theta_{\rm AI}$ correctly.
        (Right.)
        Fraction of students correctly inferring $\theta_{\rm AI}$ as a function of the number of AI-users, $k$. 
        Each dot corresponds to one simulation: pink dots indicate AI-avoiders, blue dots indicate AI-users. 
        Students who infer $\theta_{\rm AI}$ correctly and they are all AI-users except seven across all simulations.
    }
    \label{fig:larger_opponents_only}
\end{figure}

We now consider opponents-only complete networks with $n = 200$, and ask whether the findings in Section \ref{subsec: incorrect_opponent_only} still hold. 
The answer is yes: only a minority of students are able to correctly infer $\theta_{\rm AI}$, just like in Section \ref{subsec: incorrect_opponent_only}.
We consider $k = 10, 20, \ldots, 190$, and run 10 simulations  for each $k$ with randomized priors and $S_i(t)$ sequences. 
Fig.\ \ref{fig:larger_opponents_only} summarizes the results.
The left panel of Fig.\ \ref{fig:larger_opponents_only} displays $\langle \theta\rangle$ as a function of time for all AI-users and AI-avoiders and for $k=40$. 
It confirms that all students achieve asymptotic learning, but a minority (19 out of 100) infer $\theta_{\rm AI}$ correctly, and the rest do not, as in the top left panel in Fig.\ \ref{fig:opponent}. A similar outcome is observed for all the $k$ values tested. 
In the right panel of Fig.\ \ref{fig:larger_opponents_only}, each pink dot represents the number of AI-avoiders who correctly infer $\theta_{\rm AI}$ in an individual simulation divided by the total number of students, while each blue dot represents the same thing for AI-users. 
We observe that for $10 \leq k \leq 190$,  the percentage of students who correctly infer $\theta_{\rm AI}$ ranges from 1.5\% to 29\%, and all except seven turn out to be AI-users in this set of simulations.
The seven AI-avoiders who correctly infer $\theta_{\rm AI}$ occur in simulations with $k=10$. 
This agrees with the finding in \ref{subsec: incorrect_opponent_only}. 
That is, in both smaller ($n=10$) and larger ($n=200$) low-trust, opponents-only networks, the AI-users enjoy an advantage in accurately inferring $\theta_{\rm AI}$, and most AI-avoiders fail to infer $\theta_{\rm AI}$ accurately.

\subsection{Partisan student}
\begin{figure}[h!]
    \centering
    \begin{subfigure}{0.8\textwidth}        
        \includegraphics[width=\linewidth]{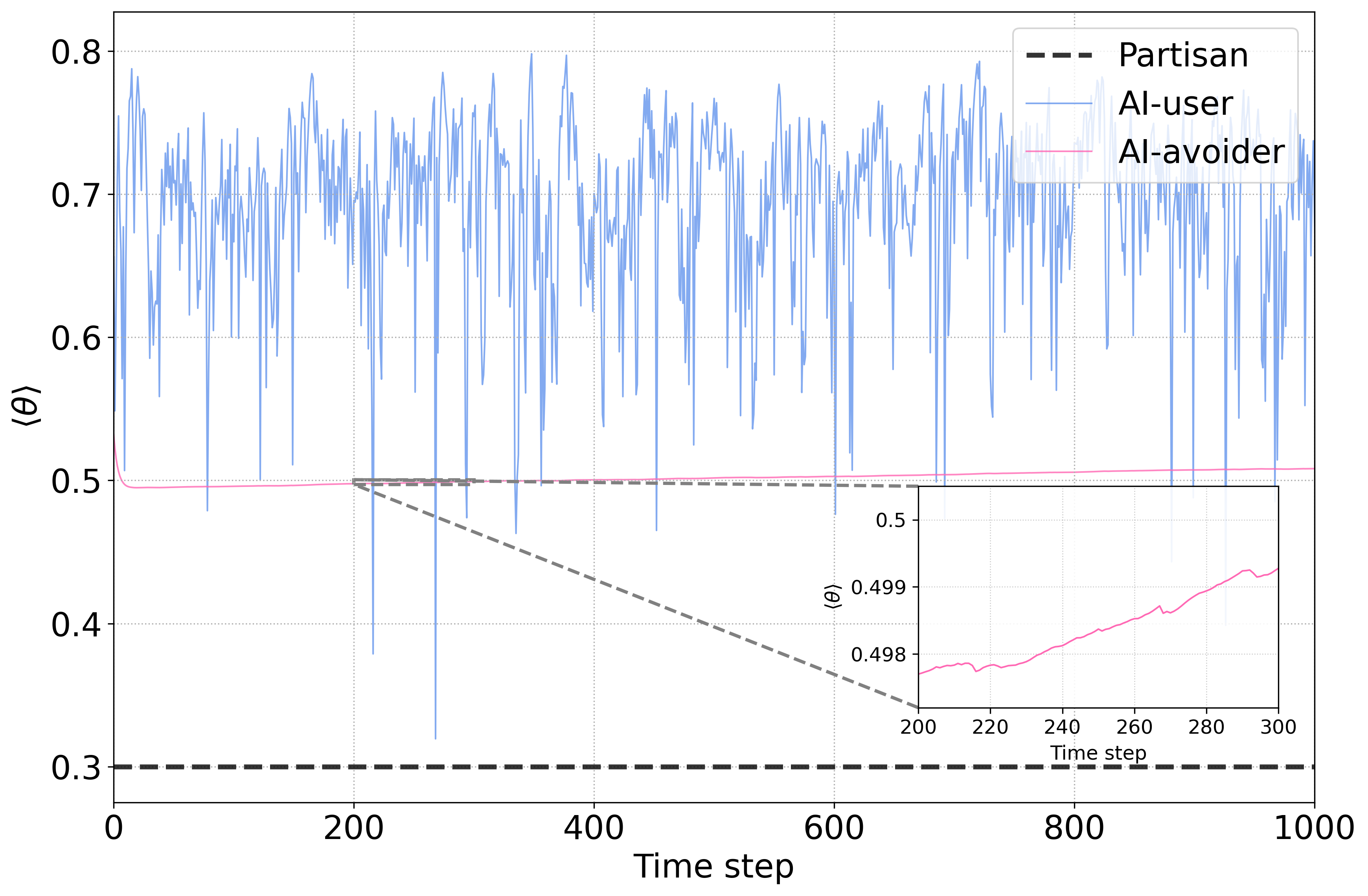}
    \end{subfigure}
    \\
    \centering
    \begin{subfigure}{0.5\textwidth}
        \includegraphics[width=\linewidth]{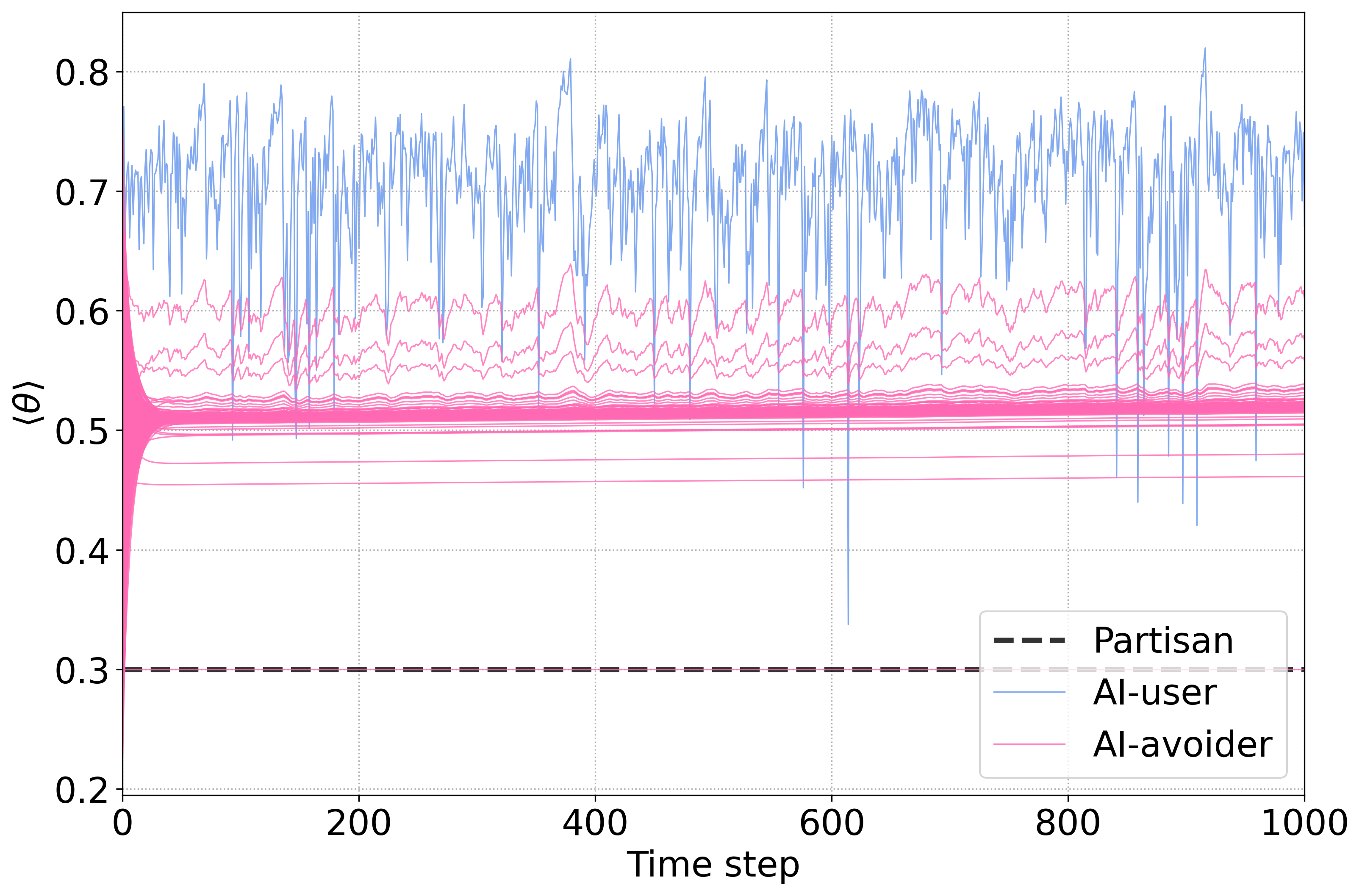}
    \end{subfigure}  
    \begin{subfigure}{0.49\textwidth}
        \includegraphics[width=\linewidth]{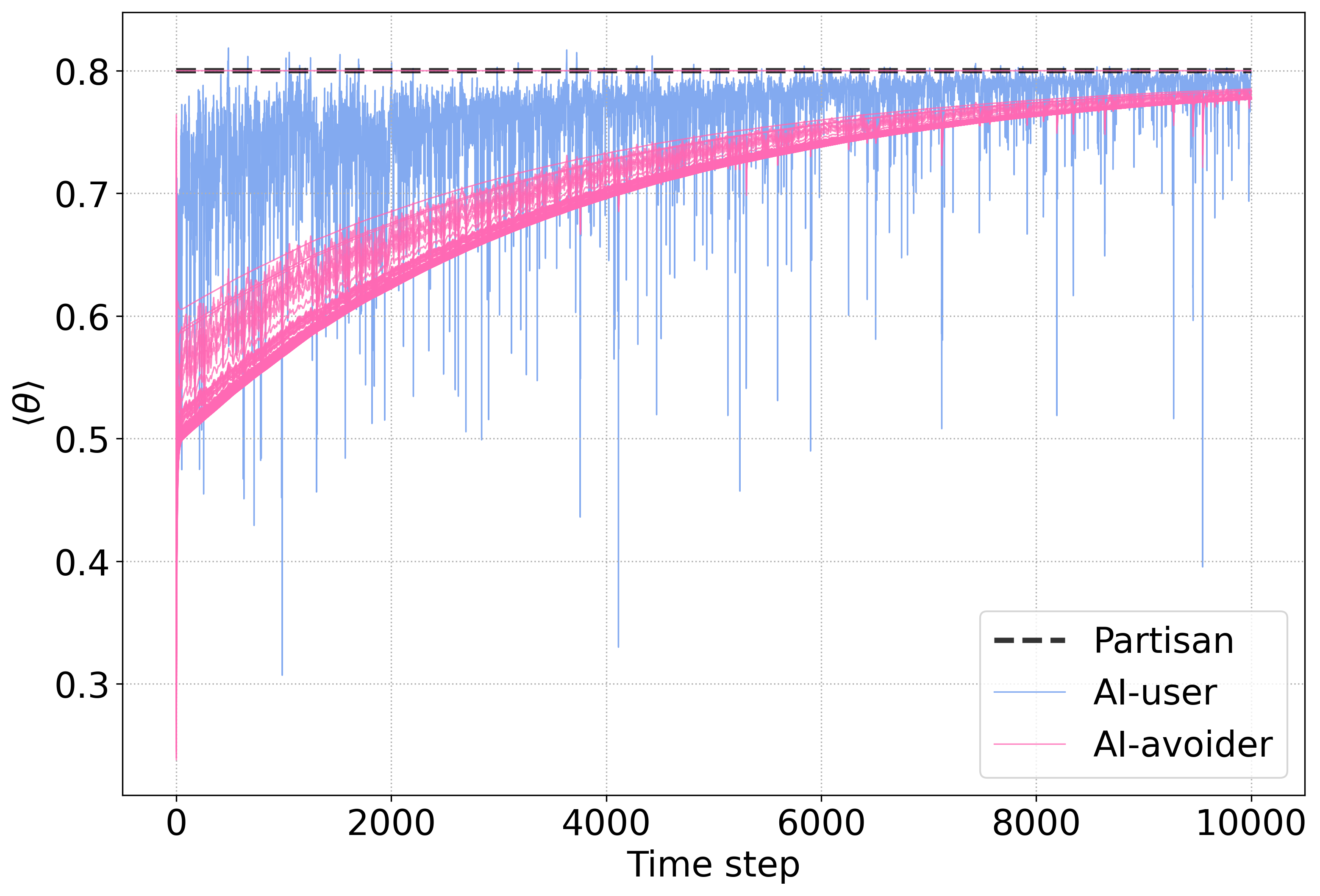}
    \end{subfigure} 

    \caption{
        Disruption by a partisan student in two larger ($n=1000$) allies-only networks: complete (top row) and Barab\'{a}si-Albert with $m=3$ (bottom row).
        The results are to be compared with the smaller allies-only networks with $n = 10$ in Fig.\ \ref{fig:partisans} (except for the bottom panel of Fig.\ \ref{fig:partisans} which portrays a mixed network).
        In all panels, the bold black dashed line indicates the partisan's belief. 
        (Top)
            Mean belief $\langle \theta \rangle$ versus time for a complete network with one AI-user, one partisan, $\theta_{\rm p} =0.3$, and $\theta_{\rm AI}=0.8$.
            The AI-user (blue curve) and a representative AI-avoider (pink curve) exhibit turbulent nonconvergence.
            The inset magnifies part of the curve for the AI-avoider.
            It confirms that the belief exhibits turbulent nonconvergence, as it fluctuates continually and nonmonotonically.
        (Bottom left.)
            Same as top panel, but for a Barab\'{a}si-Albert network with one AI-user, one partisan, $\theta_{\rm p}=0.3$, and $\theta_{\rm AI}=0.8$. 
        (Bottom right.)
            Same as the bottom left panel, but with $\theta_{\rm p} = \theta_{\rm AI}=0.8$.
        In the bottom row, it is unnecessary to distinguish the pink curves for different AI-avoiders by eye. They are plotted over one another to give a visual sense of the range of outcomes and are distinguished from the AI-user (blue curve).
        }
    \label{fig:partisans_large}
\end{figure}

We now examine how partisan students affect larger networks, replicating the experiments in Section \ref{subsec:student-partisan}.
By way of illustration, we simulate one AI-user and one partisan in allies-only complete and Barab\'{a}si-Albert networks with $n=1000$. 
The results are displayed in the top (complete) and bottom (Barab\'{a}si-Albert) rows of Fig.\ \ref{fig:partisans_large} and agree broadly with the results for $n=10$ in Fig.\ \ref{fig:partisans}. 
Specifically, for $\theta_{\rm p} = 0.3 < \theta_{\rm AI}=0.8$, AI-users and AI-avoiders exhibit turbulent nonconvergence in both network types, as observed in the top and bottom left panels of Fig.\ \ref{fig:partisans_large}. 
All AI-avoiders in the complete network achieve consensus, but some AI-avoiders in the Barab\'{a}si-Albert network do not. 
The specific behavior of each AI-avoider depends sensitively on the exact entries in $A_{ij}$, a systematic study of which lies outside the scope of this paper. 
A qualitative sense of the range of typical outcomes can be appreciated visually from the pink curves in the bottom row of Fig.\ \ref{fig:partisans_large}.
The magnified segment of the pink curve in the inset of the top panel in Fig.\ \ref{fig:partisans_large} shows that for AI-avoiders, the spread $\langle \theta \rangle_{\rm max} - \langle \theta \rangle_{\rm min}$ is smaller than for the pink curve in the top left panel in Fig. \ref{fig:partisans}. 
This aligns with Eq.\ \ref{eq:xiprimed}: the partisan’s influence on each AI-avoider is inversely proportional to the degree. 
We omit the plot for the larger complete network for $\theta_{\rm p} = 0.8 = \theta_{\rm AI}$, as it looks nearly identical to the bottom right panel of Fig.\ \ref{fig:partisans_large}, except that all AI-avoiders achieve consensus.

In smaller Barab\'{a}si-Albert networks with $n=10$, one incorrect partisan causes all students to exhibit turbulent nonconvergence, and one correct partisan leads to fast asymptotic learning, where all students infer $\theta_{\rm AI}$ correctly, as shown in \ref{sec:incomplete}.  
We now run the same experiment for a Barab\'{a}si-Albert network with $n = 1000$ and $m = 3$. 
For $\theta_{\rm p} = 0.3 \neq 0.8 = \theta_{\rm AI}$, all agents exhibit turbulent nonconvergence, although AI-avoiders no longer achieve consensus.
For $\theta_{\rm p} = 0.8 = \theta_{\rm AI}$, no AI-user or AI-avoider achieves asymptotic learning for $t\leq T$, but their mean beliefs approach  $\theta_{\rm AI}$ slowly, as seen in the bottom right panel of Fig.\ \ref{fig:partisans_large}. 
We defer longer, computationally expensive simulations to evaluate $t_{\rm a}$ to future work.

\subsection{Partisan teacher}
\label{subsec:partisan_teacher_larger}
\begin{figure}[h!]
    \centering
    \includegraphics[width=0.75\linewidth]{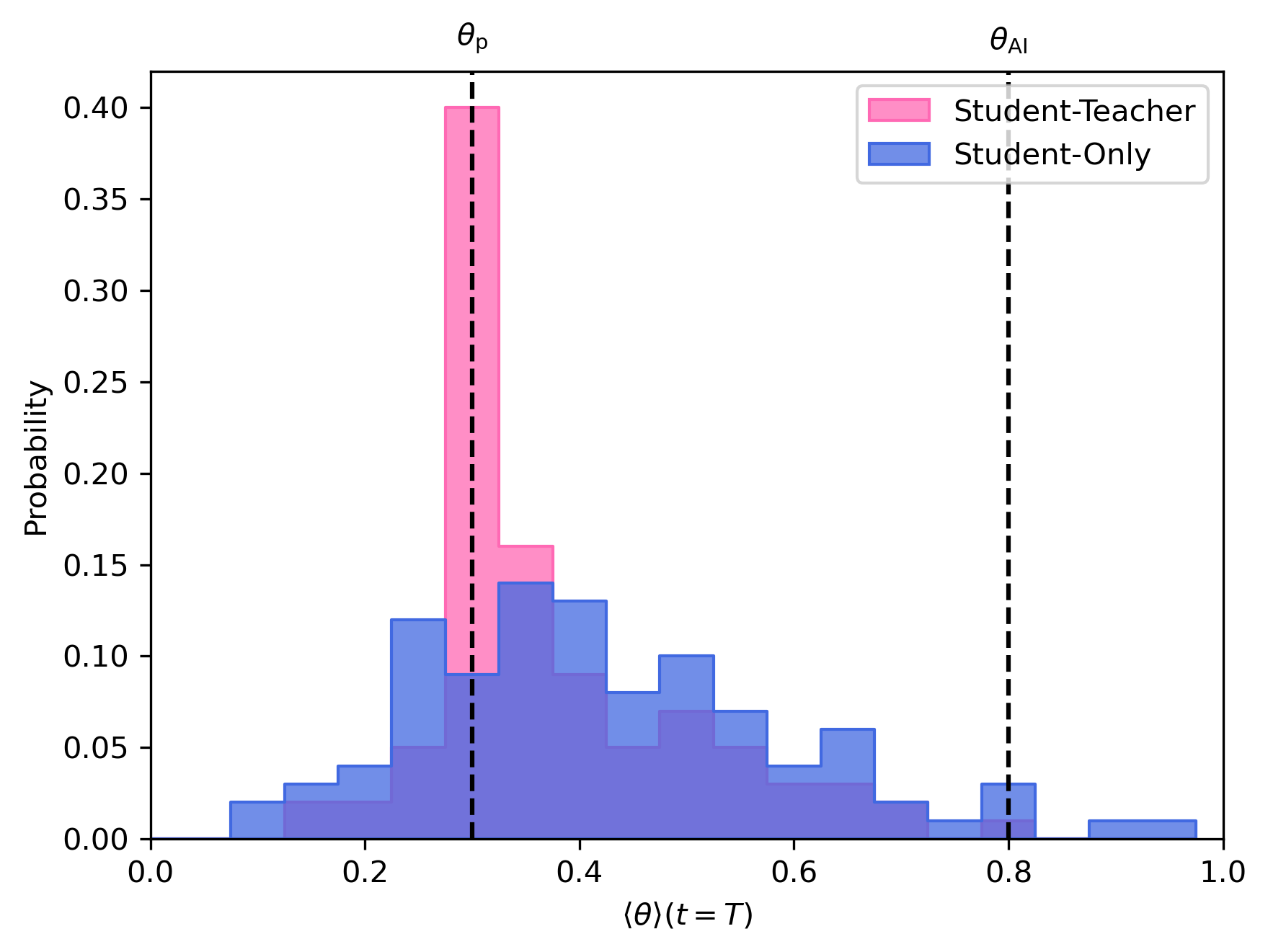}
    \caption{
        Opinion formation in larger classes: influence of a teacher in a mixed Barab\'{a}si-Albert network with 100 students, zero or one partisan teachers with $\theta_{\rm p} = 0.3 < 0.8 = \theta_{\rm AI}$, and attachment parameter $m = 3$. 
        Normalized histograms of $\langle \theta\rangle (t=T)$ are displayed for the student-only network (blue) and student-teacher network (pink).
        The black vertical dashed lines indicate $\theta_{\rm P}$ and $\theta_{\rm AI}$ respectively. 
        Students are more likely to agree with $\theta_{\rm p}$ under the influence of the partisan teacher. 
    }
    \label{fig:large_network}
\end{figure}

Finally, let us consider briefly a student-teacher network with 100 students and zero or one partisan teachers, generalizing the analysis in Section \ref{subsec:teacher-partisans} to a larger class.
The student-only subnetwork is endowed with a Barab\'{a}si-Albert structure with $n=100$, $m=3$ (i.e.\ sparse), and mixed allegiances and contains one AI-user (student 0) and 99 AI-avoiders (students 1--99) to model the interesting situation, where a small number of students are early adopters of an AI tool. The teacher, if included, is allied with all the students.
Fig. \ref{fig:large_network} presents the output of a representative simulation, where the partisan teacher is an AI-skeptic, i.e.\ $\theta_{\rm p} = 0.3 < 0.8 =  \theta_{\rm AI}$. 
We plot the histogram of $\langle\theta\rangle(t=T)$ for the 100 students, normalized to unit total probability.
The blue histogram shows the results for a student-only network, and the pink histogram shows the results when one partisan teacher is included, as above.
For both simulations, the priors, the sequence $S_i(t)$, and $A_{ij}$ are identical. 
The vertical dashed lines indicates $\theta_{\rm p}$ and $\theta_{\rm AI}$.
We find that students are more likely to believe in $\theta_{\rm p}$ under the influence of the partisan teacher ($40 \%$, versus $9 \%$ when the teacher is not included). 
The teacher is more influential in larger, sparse networks, as they have a trust relationship (here, positive) with all the students, whereas the students are sparsely connected amongst themselves.
Additionally, a minority of students correctly infer $\theta_{\rm AI}$, whether the teacher is included or not, which agrees with the result in Section \ref{subsec:teacher-partisans}.
Of course, the above experiment specializes to one particular network configuration, and should not be viewed as definitive; it is merely a guide to the sort of questions one may ask in future work.

\section{Preliminary pedagogical implications}
\label{sec:Tentative_educational_implementation}
In this section, we offer some tentative educational implications of the results in Sections \ref{sec:allies} -- \ref{sec:larger_networks}.
As noted throughout, the model in this paper is idealized, and its real-world implications should be assessed with caution. 
Nevertheless, it offers some quantitative insights into how social forces, specifically trust relationships, should be respected when deploying AI tools in an educational setting.

In a high-trust environment (allies-only network), which is desirable for multiple social and educational reasons, a single misinformed and obdurate student can disrupt the formation of accurate perceptions about $\theta_{\rm AI}$, driving the entire student cohort away from asymptotic learning (i.e.\ turbulent nonconvergence) or towards learning asymptotically the wrong $\theta_{\rm AI}$.
Hence there may be a tactical advantage for students to be a ``fast follower'' rather than an ``early adopter'', as occurs in other technological settings \cite{ankney_fast-follower_2005} or in financial markets \cite{kopel_commitment_2008}. 
In a low-trust environment (opponents-only network), a single misinformed and obdurate student is less disruptive, but there are other problems: students form stable yet erroneous perceptions about $\theta_{\rm AI}$, interfering with their ability to make sound grade- and learning-maximizing decisions. 
AI-users enjoy an advantage over AI-avoiders in an opponents-only network, raising equity concerns, but the advantage is slight. 
Overall, peer pressure coupled with widespread mistrust diminish the potential value of the AI tool for everybody, as students respond more to the flawed perceptions of their opponents than their own experiences of the tool’s actual proficiency.

Inequitable access to AI tools (e.g.\ due to socioeconomic factors) is more worrying in a high-trust environment, where AI-users have more AI-using allies, granting them more information and reinforcing their privileged position, as demonstrated by the asymptotic learning time decreasing, as the number of AI-users increases. 
In a mixed network, featuring both high- and low-trust relationships, students who are opponents of the AI-user are still able to obtain information about the AI tool though CNPs, helping students with more friends to reach the correct conclusion.
These dynamics highlight further the risk of deepening educational inequalities, where students with greater access to AI tools or more AI-using peers are more likely to succeed, while those without such access or network support are left behind.
Educators should not be troubled by the diversity of unsettled beliefs regarding $\theta_{\rm AI}$ in the classroom, as this is a natural and unavoidable result of students' priors and the network's structure. 
When a partisan teacher is introduced into the network, they function as an influential node, guiding students' opinions toward their own beliefs, regardless of accuracy. 
It is, therefore, the teacher's responsibility to independently and accurately ascertain $\theta_{\rm AI}$.

The sometimes counterintuitive role of trust networks should be considered carefully by educators when crafting policies to ensure the healthy use of AI tools in the classroom.
AI tools do not merely introduce new technology; they reshape social and academic hierarchies, interacting helpfully or unhelpfully (and usually in a complicated manner) with existing trust relationships and socio-psychological forces (peer pressure) and potentially deepening existing inequalities. 
Equal access to AI tools in a high-trust environment is a good start but it is not enough, as the results in Sections \ref{sec:allies} and \ref{sec:partisan_student} show. 
A case can be made for publishing transparent assessments of the proficiency of AI tools, which are available to all students, by a body of unbiased and nonideological experts in pedagogy and statistical analysis, who are scrupulously independent of the non- and for-profit organizations designing the AI tools themselves.

One may ask whether the results in this paper apply to educational resources beyond AI tools. 
What is the difference, for example, between an AI tool and a set of authoritative notes or a textbook recommended by a teacher? 
The answer to this interesting question depends partly on how students exploit an educational resource to address assessment tasks and, relatedly, how teachers formulate assessment tasks.
If an assessment task is mainly a test of locating and reporting known facts (``What is the chemical formula of water?'' ``H$_2$O.''), then looking up a textbook is broadly equivalent to consulting an AI tool in an oracular manner, and the opinion dynamics investigated in this paper are relevant directly. 
A good textbook has high quality $\theta_{\rm book} \approx 1$, a poor textbook has low quality $\theta_{\rm book} \approx 0$, and students who do not consult the textbook themselves must rely on other students to gauge $\theta_{\rm book}$. 
On the other hand, if an assessment task involves nuanced and creative thinking (e.g.\ solving a one-off Lagrangian mechanics problem, which does not appear on the internet), a textbook and an AI tool contribute differently to assessment outcomes and, relatedly, to the opinion dynamics investigated in this paper. 
The textbook communicates fundamental knowledge through facts, context, worked examples, and so on but it does not generate an oracular response to an assessment task. 
The grade a student receives is a function mainly of the student's own proficiency, which makes it harder to infer $\theta_{\rm book}$. 
Of course, it is possible to use an AI tool in the same way to discover and then apply the facts, context, and worked examples contained in textbooks and other sources.
However, if the AI tool is consulted in an oracular manner, the student's own proficiency does not enter the opinion dynamics, and the results in this paper are relevant directly. 
The role of trust networks in shaping perceptions about the quality of educational resources other than AI tools, and the learning and adaptation which occur through feedback from AI tools and human teachers, are interesting topics of inquiry, which we defer to future work.

\section{Conclusion}
\label{sec:conclusion}
In this paper, we develop an idealized model to evaluate how positive or negative trust relationships between students and teachers influence the emergence of perceptions about the proficiency of AI tools in educational settings. 
In the model, students form perceptions through a mix of independent observations (i.e.\ interpreting their grades when they use an AI tool) and peer pressure (i.e.\ converging towards or diverging from the perceptions of other students whom they trust or distrust respectively). 
The model accommodates partisans (obdurate agents with fixed perceptions) and a teacher (who is connected typically to every student) as well as the baseline scenario, where all participants are persuadable. 
We investigate the long-run behavior through Monte Carlo simulations and classify the results in terms of three outcomes identified previously in the literature, namely asymptotic learning (with or without consensus), turbulent nonconvergence, and intermittency \cite{bu_discerning_2023,low_discerning_2022,low_vacillating_2022}. 
We summarize the key findings in Table \ref{tab:result_tab}, organizing them according to the types of trust relationships, the presence or absence of partisans, and whether the partisans' beliefs are correct or not. 
In allies-only networks without partisans, all students achieve asymptotic learning and infer $\theta_{\rm AI}$ correctly.
AI-avoiders achieve asymptotic learning faster than AI-users.
As the number of AI-users increases, the asymptotic learning time for AI-users decreases.
Asymptotic learning is disrupted, when there are one or more partisan students with $\theta_{\rm p} \neq \theta_{\rm AI}$, whereas asymptotic learning is achieved for $\theta_{\rm p} = \theta_{\rm AI}$.
In opponents-only networks, all students achieve asymptotic learning, yet only a minority of students infer $\theta_{\rm AI}$ correctly, with AI-users reaching the correct conclusion more often.
The fraction of students who infer $\theta_{\rm AI}$ correctly depends on properties of the network and the priors; a systematic investigation is postponed to a forthcoming paper.
In mixed networks, turbulent nonconvergence and intermittency are observed, with long-term outcomes depending on the specific pattern of connections. 
Broadly and qualitatively speaking, the above results hold similarly in small and large classes with $10\leq n \leq 10^3$, as well as in complete (i.e.\ densely connected) and Barab\'{a}si-Albert (i.e.\ sparsely connected) networks.

One must exercise caution when seeking to distil policy implications applicable to real-world educational settings from a study like this. The model in this paper is idealized; it does not capture fully the nuanced psychology of individual and collective human behavior. 
Nevertheless, Monte Carlo multi-agent simulations are instructive because they reveal subtle and counterintuitive network effects --- caused by trust relationships in this paper --- which are difficult to identify by other, noncomputational means. 
Therefore, with due reserve, we offer three preliminary pedagogical and policy recommendations for consideration by educators, which may help to combat certain unintended outcomes.
\begin{itemize}
    \item Educators should be aware that trust relationships within the classroom significantly influence how students perceive and adopt AI tools.  They should talk about trust relationships and network pressures openly in the classroom, so that students are consciously aware of how they arrive at their own perceptions and become more sophisticated not only at using the technology but also at assessing its proficiency critically as an informed group. 
    \item Unequal access to AI tools deepens educational inequalities, as better-resourced students who may be connected (e.g. through friendship ties) to more AI-users enjoy an advantage in assessing accurately the proficiency of AI tools. Educators should play an active role in deepening knowledge exchange about AI tools among students by promoting class-based discussion groups (e.g.\ through an online learning management system), so as to ensure that information about the tools' proficiency is disseminated equitably, even if some students enjoy better access than others to the tools themselves.  
    More generally, students are more likely to form accurate perceptions in a higher-trust (i.e.\ allies-rich) environment. Discussion groups help to foster higher trust, as do team-building exercises like joint assessment tasks and research projects. 
    \item In tandem, educators should also disseminate unbiased, independent, scientific studies of the proficiency of relevant AI tools, when such studies exist, to every student in their class. This helps students form more accurate perceptions about proficiency, by reducing the stochasticity associated with the external signal $S_i(t)$. However, better independent information, even when disseminated equitably, is not a panacea. A key counterintuitive conclusion of this paper is that social influences, transmitted via a complicated network of trust relationships, can overwhelm an external signal under certain circumstances, even if the accuracy of the external signal is high.
\end{itemize}

\begin{sidewaystable}[h!]
    \centering
    \begin{tabular}{|>{\centering\arraybackslash}p{2.3cm}|>{\centering\arraybackslash}p{3.5cm}|>{\centering\arraybackslash}p{3cm}|>{\centering\arraybackslash}p{3cm}|>{\centering\arraybackslash}p{3cm}|>{\centering\arraybackslash}p{3cm}|}
    \hline
    & \textbf{No Partisan} & \multicolumn{2}{p{6cm}|}{\textbf{Partisan Student}} & \multicolumn{2}{p{6cm}|}{\textbf{Partisan Teacher}} \\ \hline
    \textbf{Trust relationships}                   &                      & $\theta_{\rm p} \neq \theta_{\rm AI}$ & $\theta_{\rm p} = \theta_{\rm AI}$ & $\theta_{\rm p} \neq \theta_{\rm AI}$ & $\theta_{\rm p} = \theta_{\rm AI}$ \\ \hline
    \textbf{Allies-only}   & 
        \begin{itemize}[topsep=0pt, leftmargin=*]
        \item All students reach asymptotic learning and infer $\theta_{\rm AI}$ correctly.
        \item AI-avoiders reach asymptotic learning faster than AI-users.
        \item As the number of AI-users in the network increases, the asymptotic learning times for the AI-users decrease.
        \end{itemize} & 
        Students' beliefs exhibit turbulent nonconvergence between $\theta_{\rm p}$ and $\theta_{\rm AI}$. & All students reach asymptotic learning and infer $\theta_{\rm AI}$ correctly. & Same as partisan student. & Same as partisan student. \\ \hline
    \textbf{Opponents-only} & \multicolumn{5}{p{15.8cm}|}{
        \begin{itemize}[topsep=0pt, leftmargin=*]
            \item All students reach asymptotic learning.
            \item A minority of students ($\approx 16 \%$) correctly infer $\theta_{\rm AI}$. 
            \item More AI-users than AI-avoiders reach the correct conclusion.
        \end{itemize}} \\ 
        \hline
    \textbf{Mixed}         & 
    \begin{itemize}[topsep=0pt, leftmargin=*]
        \item Turbulent nonconvergence and intermittency are observed. 
        \item Long-term behavior of students depends sensitively on the pattern of network connections. 
    \end{itemize}& 
    \multicolumn{2}{p{6cm}|}{
        \begin{itemize}[topsep=0pt, leftmargin=*]
            \item Turbulent nonconvergence and intermittency are observed. 
            \item Long-term behavior of students depends sensitively on the pattern of network connections, especially the relationship between the partisan and the AI-user.
        \end{itemize}} & 
        \begin{itemize}[topsep=0pt, leftmargin=*]
            \item The teacher is outweighed by many students who oppose the AI-user, leading the AI-user to have belief $x_{\rm AI-user}(t, \theta_{\rm p}) = 0$. 
            \item In large networks, students are more likely to infer $\theta_{\rm p}$ when a partisan teacher is in the network. 
            \item Students who oppose the AI-user obtain more information about the AI tool's proficiency when they have more CNPs.
        \end{itemize}& 
        \begin{itemize}[topsep=0pt, leftmargin=*]
            \item Students' beliefs are non-zero at multiple $\theta$ values.
        \end{itemize}\\ 
        \hline
    \end{tabular}
    \caption{Summary of key findings in terms of the network of trust relationships and the presence and beliefs of partisan students and teachers }
    \label{tab:result_tab}
\end{sidewaystable}

In this paper, we focus deliberately on the important role played by trust networks in shaping perceptions about AI tools while keeping the rest of the model idealized to simplify the interpretation of the results. 
The model can be generalized in many ways to make it more realistic. 
First and perhaps foremost, one may liberate individual students to choose whether they spurn or seek AI assistance at every time step, instead of being locked into an unchanging strategy as in this paper. 
It makes sense to introduce the latter generalization in tandem with another, related one: every student possesses a (possibly uncertain) belief about their own intrinsic proficiency, so that they only seek AI assistance when they believe the AI tool to be more proficient than they are themselves unassisted (and then only if they have no moral or other qualms). 
Implementing the two generalizations simultaneously makes it natural to add a third: students may elect to delegate part of an assessment task to the AI tool and do the rest themselves, so that their score is a mixed reflection of their and the AI tool's proficiency\footnote{
    Many partial delegation strategies can be imagined. For example, a student may divide an assessment task into disjoint pieces and delegate some to the AI tool while doing the others themselves, or they may delegate the whole assessment task to the AI tool for a first pass and refine the AI output themselves.}. 
Liberating usage strategies to co-evolve with perceptions about proficiency is likely to produce complicated and less intuitive  dynamics more often, such as turbulent nonconvergence and intermittency \cite{bu_discerning_2023,low_vacillating_2022}. 
Second, we assume here that all agents infer the proficiency of a single AI tool, overlooking the reality that multiple tools exist with different proficiencies.
Some students may not have access to the latest technology, e.g.\ tools built with the most advanced LLMs, due to financial and other barriers, thereby introducing disparities in educational outcomes.
Finally, it is important to validate theoretical predictions against real-world data. 
This is done best by collaborating with social scientists, who have experience in designing and implementing experiments with human participants, as well as in navigating the complexities of human data collection and interpretation. 
Such empirical studies are resource-intensive and fall outside the scope of this theoretical paper but they are an important avenue for future work.

\section*{Acknowledgements}

AM acknowledges funding from the Australian Research Council Centre of Excellence for Gravitational Wave Discovery (OzGrav) (CE230100016). 
YB is supported by a Research Training Program Scholarship from the Australian Commonwealth Government and the University of Melbourne.

\appendix
\section{Partisan students in incomplete networks}
\label{sec:incomplete}
In this appendix, we repeat the simulations in Section \ref{subsec:student-partisan} for incomplete networks.
Specifically, by way of illustration, we study an allies-only, Barab\'{a}si-Albert network with $n=10$ and $m=3$.
The goal is to test if the results in Section \ref{subsec:student-partisan} hold unchanged for incomplete networks.
The answer, broadly and qualitatively speaking, is yes, as we now verify.

\begin{figure}[h!]
    \begin{subfigure}{0.5\textwidth}
        \includegraphics[width=\linewidth]{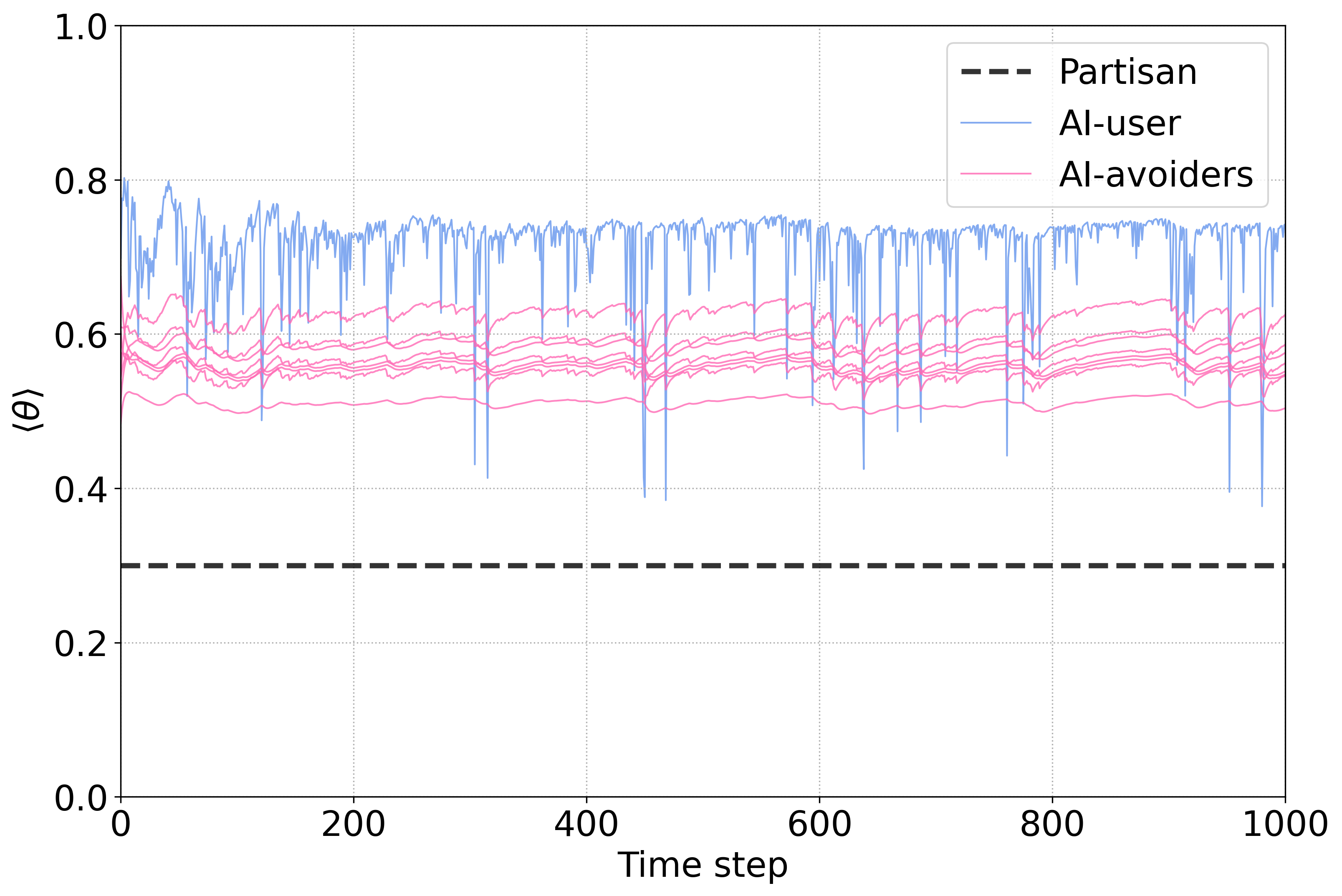}
    \end{subfigure}
    \hspace*{\fill}
    \begin{subfigure}{0.5\textwidth}
        \includegraphics[width=\linewidth]{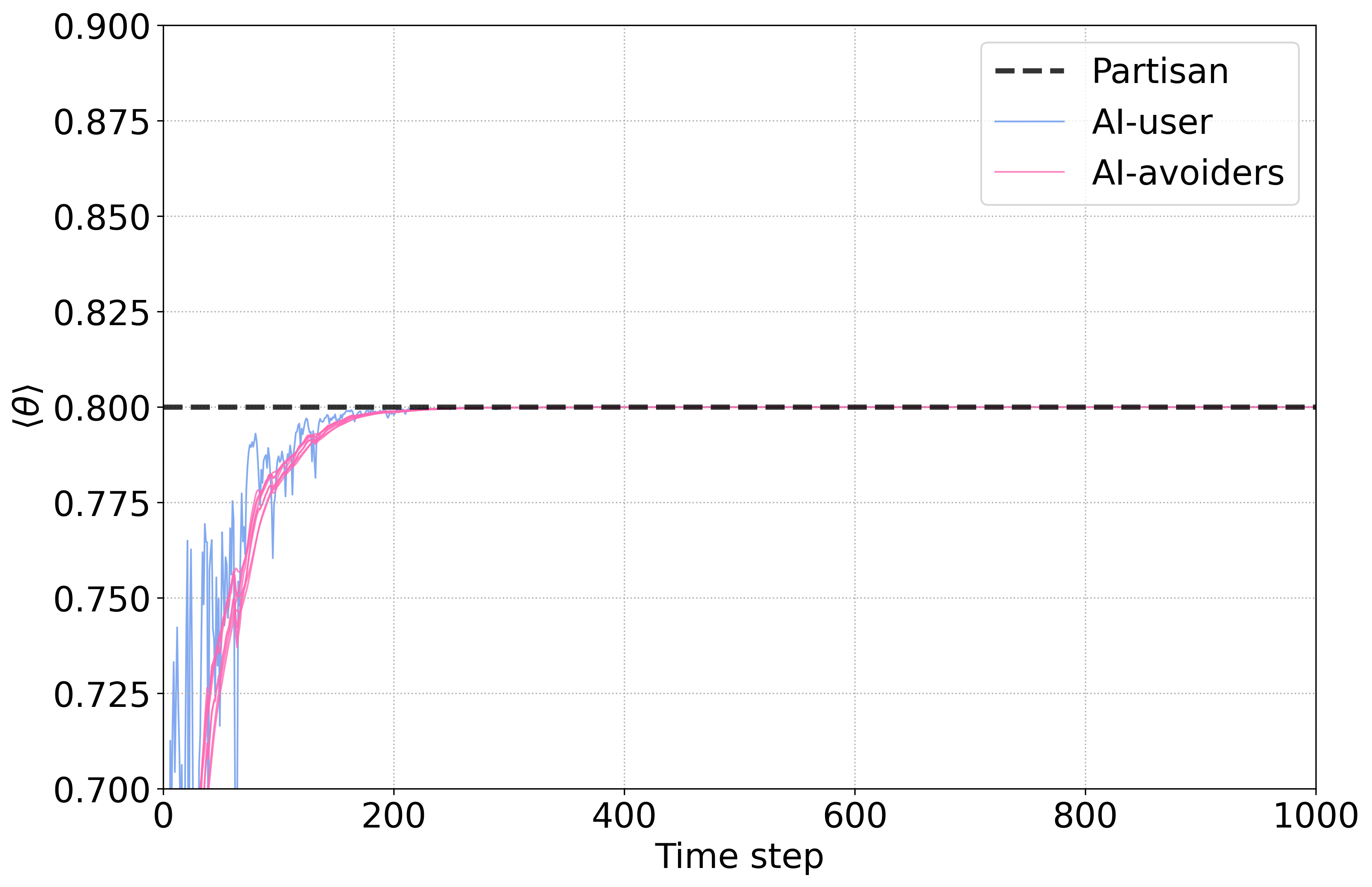}
    \end{subfigure}%
    \caption{
        Disruption by a partisan student in an allies-only Barab\'{a}si-Albert network with $n=10$ and $m=3$.
        (Left.)
        Mean belief $\langle \theta \rangle$ as a function of time with one AI-user (blue curve), eight AI-avoiders (pink curves), one partisan (black dashed curve), $\theta_{\rm p} = 0.3$, and $\theta_{\rm AI} = 0.8$. Both the AI-user and AI-avoiders exhibit turbulent nonconvergence, agreeing with the result in allies-only complete networks (see Section \ref{subsec:student-partisan}). 
        (Right.)
        Same as the left panel, but with $\theta_{\rm p} = \theta_{\rm AI} = 0.8$.
        As in complete networks, the AI-user and AI-avoiders achieve asymptotic learning.
        }
    \label{fig:BA_partisans}
\end{figure}

First, let us consider one partisan with $\theta_{\rm p} = 0.3$.
We find that turbulent nonconvergence is still observed for all students. 
We plot the mean belief $\langle \theta \rangle$ of the partisan (black dashed curve), the AI-user (blue curve), and AI-avoiders (pink curves) versus time in the left panel of Fig. \ref{fig:BA_partisans} for a representative simulation. 
As with complete networks, AI-avoiders evolve to have $x_i(t, \theta = \theta_{\rm AI}) \neq 0$ and $x_i(t, \theta = \theta_{\rm p}) \neq 0$.
AI-avoiders who are directly connected to the partisan often have greater $x_i(t, \theta = \theta_{\rm p})$ than AI-avoiders who are connected to the partisans though a common ally.
The main difference with complete networks, studied in Section \ref{subsec:student-partisan}, is that AI-avoiders do not reach consensus among themselves, due to their different pattern of connections.

Next we test one partisan with $\theta_{\rm p} = 0.8 = \theta_{\rm AI}$ in a Barab\'{a}si-Albert network (again with $n = 10$, $m = 3$).
The mean belief $\langle \theta \rangle$ is plotted versus time in the right panel of Fig.\ \ref{fig:BA_partisans} for a representative simulation. 
All students achieve asymptotic learning, as observed in complete networks in Section \ref{subsec:student-partisan}. 
In 100 simulations with $k = 1$, the $t_{\rm a}$ histograms peak at $t_{\rm a} \approx 56$ and $t_{\rm a} \approx 108$ for AI-avoiders and AI-users respectively.

\newpage
\bibliographystyle{elsarticle-num} 
\bibliography{ai,OD}

\end{document}